\documentclass[10pt,english,aps,prb,twocolumn,groupedaddress,superscriptaddress,floatfix]{revtex4-2}
\usepackage[T1]{fontenc}
\usepackage[latin9]{inputenc}
\usepackage{color}
\usepackage{ql}
\usepackage{graphicx}
\usepackage{xcolor}
\usepackage{longtable}
\usepackage{babel}
\usepackage{amsmath}
\usepackage{pgffor}
\usepackage{upgreek}
\usepackage{amssymb}
\usepackage{mdframed}
\usepackage{dcolumn}
\usepackage{supertabular}
\usepackage{array}
\usepackage[unicode=true,
 bookmarks=false,  
 breaklinks=false,
 pdfborder={0 0 0},
 pdfborderstyle={},
 backref=false,
 colorlinks=true,
 linkcolor=blue]{hyperref}
\AtBeginDocument{}
\AtBeginDocument{}

\newcommand{\alphaFine}{\textcolor{black} \alpha\!\!\!\raisebox{0.5mm}{.}\,}

\begin{document}

\title{Refining spectroscopic calculations for trivalent lanthanide ions: \\ a revised parametric Hamiltonian and open-source solution
}

\author{Juan-David Lizarazo-Ferro}
\altaffiliation{These authors contributed equally to this work.}
\affiliation{School of Engineering and Department of Physics, Brown University, Providence, RI 02912, USA}
\author{Tharnier O. Puel}
\altaffiliation{These authors contributed equally to this work.}
\affiliation{Department of Physics and Astronomy, University of Iowa, IA 52242, USA}
\author{Michael E. Flatt\'e}
\email[Contact author: ]{michaelflatte@quantumsci.net}
\affiliation{Department of Physics and Astronomy, University of Iowa, IA 52242, USA}
\affiliation{Department of Applied Physics, Eindhoven University of Technology, Eindhoven, The Netherlands}
\author{Rashid Zia}
\email[Contact author: ]{rashid_zia@brown.edu}
\affiliation{School of Engineering and Department of Physics, Brown University, Providence, RI 02912, USA}

\begin{abstract}
The historical calculation of spectroscopic properties for trivalent lanthanide ions is a complex multistep process that has been prone to inaccuracies.
In this work, we revise the parametric semi-empirical Hamiltonian and address long-standing discrepancies in the literature.
We also resurface the distinctions between orthogonal and non-orthogonal operators and use orthogonal operators to provide an alternative parametric description.
Based on experimental data available in the literature, a new set of parameter values for the ions in \LaF and \liyorite is presented.
Additionally, we provide calculations of spontaneous emission rates and oscillator strengths for magnetic dipole transitions in the \LaF crystal host.
To ensure the replicability of our findings, we make available the open-source code \qlanth, accompanied by a comprehensive set of electronic files to serve as an updated reference for future calculations.
\end{abstract}
\maketitle

\section{Introduction}

Trivalent lanthanide ions ($\text{Ln}^{3+}$) are interesting because of their narrow optical transitions, which arise from the shielding of their $4f$ electrons by the outer $5s^2$ and $5p^6$ closed shells \cite{Vleck-1937,wybourne1965spectroscopic,liu2005spectroscopic}. 
Having unpaired $4f$ electrons, the $\text{Ln}^{3+}$ ions constitute spin defects useful for lasers and amplifiers \cite{deloach_evaluation_1993,jorgensen1977lasers}, magnetic resonance imaging \cite{wei2022rare}, and quantum technology \cite{atature_material_2018,awschalom_quantum_2018,PhysRevB.105.224106,uysal2024}. 
Furthermore, the distinctive intra-$4f$ transitions are appealing for exploring fundamental topics in atomic \cite{rudzikas_theoretical_2007} and optical \cite{ taminiau_quantifying_2012,li_probing_2018,horvath2023strongpurcell} physics.

The trivalent lanthanides are often theoretically described by a Hamiltonian with adjustable parameters that are fitted empirically to experimental data. 
Using this description, one obtains an approximate electronic structure, including eigenstates, from which one may calculate optical transition rates and oscillator strengths for magnetic dipole transitions \cite{dodson_2012}, forced electric dipole transitions through Judd-Ofelt theory \cite{ofelt_intensities_1962,HEHLEN2013221}, as well as g-tensors for ground-state excitations \cite{WELLS201530,Ourari-Horvath-Jeff-2023}.
Shielding of the $4f$ shell simplifies the study of a lanthanide ion across different hosts, and within the same host, systematic studies across the lanthanide row reveal that many model parameters follow relatively simple trends  \cite{Judd_Crosswhite_84,carnall_systematic_1989,NOVAK2014414}.
The canonical example of such systematic studies is the work by \bill, which has been widely cited and continues to be used to various degrees of depth.
In the simplest usage, the energy spectra for the original lanthanun trifluoride (\LaF) crystal host are simply referenced to identify observed transitions\cite{binnemans_interpretation_2015}.
In a more elaborate usage, the parameters for \LaF may be used as starting values for fitting the data for different crystal hosts \cite{cheng_crystal-field_2016, gorller-walrand_rationalization_1996}. 
Furthermore, the eigenvectors can be used to calculate the magnetic susceptibility as a function of temperature  \cite{soderholm_crystal-field_1991,loong_ground-state_1993}. 
The canonical work of \carnall is often taken as a theoretical starting point for the inclusion of additional terms to the \hamilton such as correlation crystal field \cite{li_correlation-crystal-field_1990,gruber_energy_1993}, hyperfine interaction \cite{mothkuri2021electron}, spin-correlated crystal field \cite{gruber1990comparative}, and exchange interactions  \cite{boldyrev2021thermal}.

However, the canonical approach has some shortcomings. For instance, calculations with the \hamilton are convoluted, and this complexity makes it prone to inaccuracies \cite{chen_few_2008,judd_three-electron_1994}.
The calculations generally rely on the application of the methods developed by Racah \cite{PhysRev.61.186,PhysRev.62.438,PhysRev.63.367,PhysRev.76.1352} combined with pre-calculated spectroscopic tables that have been found to contain multiple errors \cite{HANSEN19961,judd_intra-atomic_1968,chen_few_2008}.
Furthermore, the calculations are somewhat ambiguous, because there are several possible ways to parametrize the relevant interactions \cite{Judd_Crosswhite_84,judd_complete_1984}.

Even after these errors were recognized, the parameters provided by \bill more than three decades ago have been used repeatedly and without update.
For example, just recently, they have been used to estimate crystal field parameters for $\text{Dy}^{3+}$ in rare-earth oxides \cite{PhysRevB.109.054434}, 
hyperfine interactions for $\text{Ho}^{3+}$ in yttrium orthosilicate \cite{mothkuri2021electron},
dipolar interactions between $\text{Gd}^{3+}$ in rare-earth oxides \cite{PhysRevB.105.014425}, and the g-tensor for $\text{Er}^{3+}$ implanted into silica \cite{PhysRevApplied.16.034006}. 
In the end, errors in the original calculation may be compensated when fitting the model parameters, yielding similar closeness to the experimental energies, and thus obscuring the innacuracy in the eigenvectors.

This paper revisits the canonical work of \bill.
The relevant spectroscopic tables for the semi-empirical Hamiltonian are recalculated, addressing both previously noted errors as well as newly identified ones.
Then, by detailed analysis of the magnetic contributions in the calculated spectra, we present and correct for incongruencies in the claimed inclusion by \carnall of the spin-spin interaction. 
Furthermore, an ongoing challenge in physics is the difficulty of reproducing published research \cite{baker_1500_2016}.
Besides some theoretical challenges, the semi-empirical approach suffers from practical challenges, many of which we aim to resolve here.
Currently available computer codes for doing these calculations are not only outdated, but they yield incompatible results (i.e., the same set of parameters leads to distinct spectra), and their internal calculations are not sufficiently documented.
Here we present a modern, well documented, publicly available implementation written in \textit{Wolfram} language, called \qlanth~\cite{qlanth}, offering solutions to many of these deficiencies.

In Secs. \ref{sec: hamiltonian} and \ref{sec: parametric Hamiltonian}, we introduce the Hamiltonian and its mapping to a parametric semi-empirical Hamiltonian within the single-configuration approximation.
In Sec. \ref{sec: errors}, we compare our calculations to the literature, as well as to other code implementations.
Sections \ref{sec: updated level structure} and \ref{sec: magnetic dipole transitions} offer updated fitted parameters -- together with a collection of reproducible data including energies, eigenvectors, and optical transitions -- that provide an updated reference for the spectroscopy of trivalent lanthanide ions.

\section{Hamiltonian}
\label{sec: hamiltonian}

A fundamental description of the electromagnetic interactions for the $f$-electrons of lanthanide ions embedded in a crystal host may be represented by the following Hamiltonian:
\begin{align}
    {\ham} =& {\ham}_{\text{K}}+{\ham}_{\text{n}}+{\ham}_{e\textrm{-}e}+{\ham}_{\text{s-o}}\nonumber \\
     &      +{\ham}_{\text{s-s}}+{\ham}_{\text{o-o}}+{\ham}_{\text{s-o-o}} + {\ham}_\text{CF}.
     \label{eq: Hamiltonian contributions}
\end{align}
The main contributions are the kinetic (K), nuclear (n), electrostatic ($e\text{-}e$), and spin-orbit (s-o) terms. These in turn are given in operator form as:
\begin{align}
    {\ham}_{\text{K}}   &=-\dfrac{1}{2}\sum_{i}\nabla_{i}^{2}, \label{eq:kinetic}\\
    {\ham}_{\text{n}}   &\approx-\sum_{i}\dfrac{ Z_\text{eff}}{r_{i}}, \\
    {\ham}_{e\text{-}e} &= \sum_{i>j}\dfrac{1}{r_{ij}},\\
    {\ham}_{\text{s-o}}  &= \dfrac{\alphaFine^2}{2}\sum_{i} \left( \dfrac{1}{r_i} \dfrac{\partial V}{\partial r_i} \right)(\boldsymbol{s}_{i}\cdot\boldsymbol{l}_{i}).
\end{align}
In the set of equations above, $Z_\text{eff}$ is the effective atomic number as screened by the closed electron shells;
$r_i$ is the distance of the $i$-th electron to the point-charge nucleus;
$r_{ij}$ is the distance between two electrons;
$\alphaFine$ is the fine-structure constant;
$V$ is the effective central potential in which the valence electrons move; 
$\boldsymbol{s}_i$ is the electron spin;
and $\boldsymbol{l}_i$ is the electron angular momentum.
Besides the main contributions, there are additional magnetic interactions that go beyond the single-electron spin-orbit ($\boldsymbol{s}_i\cdot\boldsymbol{l}_i$) term which need to be considered to take into account relativistic corrections that are not negligible in heavier atoms.
These additional magnetic terms are particularly important for the partially filled $f$-shell in the lanthanide ions because of the large orbital angular momentum and localization of the $4f$-electrons.
These relativistic contributions may be derived from the Breit Hamiltonian resulting in two-body interactions including spin-spin $(\boldsymbol{s}_i \cdot \boldsymbol{s}_j)$, spin-other-orbit $(\boldsymbol{s}_i \cdot \boldsymbol{l}_j)$ \cite{bethe1957quantum,doi:10.1098/rspa.1962.0207,wybourne1965spectroscopic,Wybourne1970book,MJones_1971,bransden2003physics,drake2006springer}, and orbit-orbit $(\boldsymbol{l}_i \cdot \boldsymbol{l}_j)$ \cite{PhysRev.92.1448,19551029,wybourne1965spectroscopic,Wybourne1970book,drake2006springer}. These interactions can be expressed as:
\begin{align}
    {\ham}_{\text{s-s}} &= \alphaFine^{2}
        \sum_{i>j}
            \left[\frac{(\boldsymbol{s}_{i}\cdot\boldsymbol{s}_{j})}{r_{ij}^{3}}-\frac{3(\boldsymbol{r}_{ij}\cdot\boldsymbol{s}_{i})(\boldsymbol{r}_{ij}\cdot\boldsymbol{s}_{j})}{r_{ij}^{5}}\right],\\
    {\ham}_{\text{o-o}} &= -\frac{\alphaFine^{2}}{2} \sum_{i<j} \left[\frac{(\boldsymbol{p}_{i}\cdot\boldsymbol{p}_{j})}{r_{ij}}+\frac{\left( \boldsymbol{r}_{ij}\cdot(\boldsymbol{r}_{ij}\cdot\boldsymbol{p}_{i})\boldsymbol{p}_{j}\right)}{r_{ij}^{3} } \right], \\
    {\ham}_{\text{s-o-o}} &= -\alphaFine^{2}\sum_{i>j}\left(\frac{\boldsymbol{r}_{ij}}{r^3_{ij}}\times\boldsymbol{p}_{i}\right) \cdot(\boldsymbol{s}_{i}+2\boldsymbol{s}_{j}),\label{eq:soo}
\end{align}
where $\boldsymbol{p}_i$ is the electron momentum.
Finally, the action of the host is included as an effective  potential with a symmetry appropriate for the crystal host. 
In its original inception \cite{bethe1929termaufspaltung}, the crystal field was understood to have a purely electrostatic origin, such that a charge density $\rho(\boldsymbol{R})$ has an associated potential given by:
\begin{align}
    V_\text{CF} (\boldsymbol{r}_i) = \int\frac{\rho(\boldsymbol{R})}{|\boldsymbol{r}_i - \boldsymbol{R}|} d \boldsymbol{R},
\end{align}
such that $\ham_\text{CF} = \sum_i (-e) V_\text{CF}(\boldsymbol{r}_i)$.
Presently, the electrostatic interpretation of the crystal field is not considered to be accurate \cite{newman2000crystal}.
However, the symmetry arguments that lead to its mathematical form as an expansion in spherical harmonics are still valid and are independent of the electrostatic interpretation.

\section{Parametric Semi-empirical Hamiltonian for the $4f$-electrons in the single-configuration description}
\label{sec: parametric Hamiltonian}

An approximate solution to the Hamiltonian in Eq. (\ref{eq: Hamiltonian contributions}) can be obtained by limiting the Hilbert space to the $4f^N$ ground configuration, where $N$ is the number of electrons in the $4f$ shell.
The single (ground) configuration $4f^N$ opposes a more generic approach that includes the occupation of additional (excited) orbitals in the Hilbert space, discussed later in the configuration-interaction treatment.
In the single configuration description, a particular basis state is labeled by specifying its $S$, $L$, $J$, and $M_J$ quantum numbers, and is usually written as $| \psi \rangle = \left| f^N \tau SLJM_J \right>$ state, where $\tau$ is an additional quantum number that accounts for the degeneracy of the LS terms.
The matrix elements are analyzed using the algebra of irreducible tensor operators developed by Racah~\cite{PhysRev.61.186,PhysRev.62.438,PhysRev.63.367,PhysRev.76.1352}, in which the angular part can be calculated explicitly, and the radial part subsumed as multiplicative coefficients.
More precisely, matrix elements of the form
$
\left\langle \psi_{i}\right|{\ham}\left|\psi_{j}\right\rangle
$ need to be determined,
where $i,j$ run over all the states $\left| f^N \tau SLJM_J \right>$. 
Racah algebra allows computing each matrix element in the form 
$
p \left\langle \psi_{i}\right| \hat{\cal O} \left|\psi_{j}\right\rangle,
$
where the operator $\hat{\cal O}$ contains the angular part and the radial part $p$ is a coefficient later treated as a fitting parameter.
Once these reduced matrix elements have been determined for the configurations $f^N$ with $N\leq 3$, then the method of fractional parentage~\cite{PhysRev.76.1352} is used to determine them for $f^{N+1}$.
Further details of the calculations of such matrix elements is well documented in Refs. \cite{wybourne1965spectroscopic,cowan1981theory} and will not be repeated here.
In the following, however, we will summarize how the form of Hamiltonian shown in Eq. (\ref{eq: Hamiltonian contributions}) is transformed to the form actually used in calculations as given ahead in Eq. (\ref{eqn:hamiltonian}).
A summary of this mapping is given in Appendix \ref{sec: mapping}.

First, the kinetic energy and nuclear potential energy ($\ham_K$ and $\ham_n$) depend only on the number of electrons $N$, are  constant over the states of the ground configuration, and are noted as $\ham_0$ in Eq. (\ref{eqn:hamiltonian}).
Experimentally we have access only to energy differences rather than absolute values, as such this constant energy shift is disregarded.
Second, the next largest contribution to the energy spectrum comes from the Coulomb term ($\ham_{e\text{-}e}$).
The calculation of the matrix elements involving the Coulomb term leads to the set of parameters $\Fk{k}$ and operators $\op{f}_k$, where $k$ corresponds to the  order of the Slater radial integral, and where only $k=0,2,4,6$ need to be considered as per selection rules of integrals of products of three spherical harmonics.
Third, a large contribution also comes from the spin-orbit term ($\ham_\text{s-o}$), which reduces to a single fitting parameter $\spinZeta$ and the operator $\op{s}_i \cdot \op{l}_i$ for the $i$-th electron.

\subsection{Configuration, magnetic, and crystal field interactions}

The single-configuration approximation can be much improved by the inclusion of effective terms originating from configurations other than $4f^N$.
These effects constitute the next most important correction to the Hamiltonian and may be calculated using  perturbation theory without increasing the size of the Hilbert space.
For example, the configuration interaction analysis relevant to the ground configuration $4f^3$ of  $\text{Nd}^{3+}$ considers configurations like $4f^2 6s$, $4f^2 5d$, and $4f^1 6s^2$.
Luckily, for the trivalent lanthanides, the center of the energy spectrum of the $4f$-configuration is well isolated ($\sim 10^4 \text{ cm}^{-1}$) from other configurations~\cite{Dieke:61}, thus interacting only via high energies. 
This justifies treating the interacting configurations as  perturbations to the single-configuration description, which to second order adds terms like $\left<\psi\right| G \left|m\right> \left<m\right| G \left|\psi'\right>/\Delta E_m$ to the Hamiltonian~\cite{PhysRev.132.280}, where $\left|m\right>$ represents a configuration other than $4f$ and $G$ the Coulomb operator. 
These perturbations are responsible for the parameters $\casimirAlpha$, $\casimirBeta$, and $\casimirGamma$ in Eq. (\ref{eqn:hamiltonian}), as well as a screening of the $\Fk{k}$ parameters.
Moreover, still within second-order-perturbation theory, the case of one-electron excitation either from or into the $4f^N$ shell (e.g. $4f^2 5d$ in $\text{Nd}^{3+}$) needs an additional set of parameters $\Tk{k}$, in which the orbital part involves coordinates of three electrons and thus represents an effective three-body interaction~\cite{PhysRev.132.280,PhysRev.141.4}.
Terms of third order from perturbation theory lead to additional parameters $T^{(11)}, \ldots , T^{(19)}$ with three-body type operators ~\cite{PhysRev.132.280,Judd_Suskin_84}, which will be disregarded here following common practice.

The configuration interaction analysis can yield additional terms in which different operators are correlated to one another.
In addition to the previously discussed contributions, second-order perturbation theory brings extra terms, like $\left<\psi\right| \Lambda \left|m\right> \left<m\right| G \left|\psi'\right>/\Delta E_m$, where $\Lambda$ is the spin-orbit operator, see Refs.~[\onlinecite{PhysRev.134.A596,judd_intra-atomic_1968}].
That term is called electrostatically-correlated-spin-orbit interaction, because of its mixing of $G$ and $\Lambda$. 
In Racah formulation, this term contributes similarly to the spin-other-orbit discussed below, but also requires a set of independent parameters $\Pk{k}$ called the pseudo-magnetic parameters.

The magnetic dipolar interaction ($\ham_\text{s-s}$) contributes with the parameters $\Mk{k}$, which have a radial integral form as defined by Marvin~\cite{PhysRev.71.102}, and appears with the $\hat{m}_k^\text{s-s}$ operators in the Hamiltonian.
The effects of the spin-spin contribution in the energy spectrum (on Pr, Nd, Er in LaCl$_3$) is discussed in Ref. \cite{yeung_new_2013}.
The next magnetic interaction, the spin-other-orbit ($\ham_\text{s-o-o}$), contributes in two manners. 
First, it contributes to modifying the spin-orbit parameter $\spinZeta$. 
Second, it also requires $\Mk{k}$ as fitting parameters. 
Although sharing the same fitting parameters, we explicitly write the operator $\hat{m}_k^\text{s-o-o}$ for the spin-other-orbit in Eq. (\ref{eqn:hamiltonian}) to distinguish it from the dipolar interaction. 
The reason for that will become clear in the next section, where we make a quantitative comparison with literature results.
The orbit-orbit interaction ($\ham_\text{o-o}$) contributes similarly as $\casimirAlpha$, $\casimirBeta$, and $\casimirGamma$ and therefore is parametrically absorbed in the fitting.

Finally, the effect of the crystal field ($V_\text{CF}$) from the host material on the $4f$-electrons of the lanthanide ions is taken into account with the ${\cal B}_q^{(k)}$ parameters. 
The non-zero crystal field parameters are determined by the point symmetry of the location of the ions inside of the host crystal, where, less symmetry requires more parameters (see discussion in Ref. \cite{binnemans_interpretation_2015}).
The calligraphic $\Bcomplexkq{k}{q}$ carries both the real and imaginary coefficients, such that  $\Bcomplexkq{k}{q} = \Bkq{k}{q} + i\Skq{k}{q}$.

\begin{table}[!htbp]
    \begin{center}
        \begin{tabular}{|c|c|}
            \hline
            Parameters & Splitting Range ($\invcm$) \\ 
            \hline
            $\Fk{k}$ & $(20000 - 40000)$ \\
            $\spinZeta$ & $(2000 - 10000)$ \\
            $\casimirAlpha, \casimirBeta, \casimirGamma$ & $(300 - 1200)$ \\
            $\Tk{k}$ & $(100-2000)$ \\
            $\Pk{k}$ & $(20-200)$ \\
            $\Mk{k}$ & $(10-40)$ \\
            $\Bkq{k}{q}$ & $(40 - 500)$ \\
            \hline
        \end{tabular}
    \end{center}
    \caption{Approximate ranges of energy splittings contributing to the \hamilton in the case of \LaF.}
    \label{table:orders-of-mag}
\end{table}

\subsection{Parametric Semi-empirical Hamiltonian}

In conclusion, the form of the parametric semi-empirical Hamiltonian is
\begin{align}
\ham_\text{para} &=  \ham_{0}
 + \sum_{k=0,2,4,6} \Fk{k} \hat{f}_{k}
 + \spinZeta \sum_{i=1}^N  \paren{ \hat{s}_i \cdot \hat{l}_i} \nonumber \\
 &\,\,  + \casimirAlpha \hat{L}^2 
  + \casimirBeta \casimir{\Gtwo}
   + \casimirGamma \casimir{\SO{7}} + \hamEffectiveThreeBody \nonumber \\
 &\,\,\,\,\,\, 
 + \hamSOOplusECSO \nonumber \\
 &\,\,\,\,\,\,\,\,  + \hamCrystalFieldALT,
\label{eqn:hamiltonian}
\end{align}
where $\hat{L}$ is the total angular momentum of all $4f^N$ electrons, $\hat{\cal C}(\Gtwo)$ and $\hat{\cal C}(\SO{7})$ are the Casimir operators of groups $\Gtwo$ and $\SO{7}$, respectively. $\hat{C}^{(k)}_q$ are spherical harmonics with the Racah normalization convention. 
This parametric Hamiltonian is then considered semi-empirical when its parameters are found by fitting experimental data. 
On first approximation, many parameters in the \hamilton have simple integral expressions in terms of the radial part of the single-electron orbitals. However, these expressions become significantly more complex when configuration interaction is included, and estimating the single-electron orbitals is itself a non-trivial task. The semi-empirical approach addresses these theoretical challenges by directly referencing experimental data to determine the value of the model's parameters.

To quantify the relative contribution of each term in Eq. (\ref{eqn:hamiltonian}) for the case of lanthanide ions in \LaF, we looked at the maximum energy splitting generated by each term individually. We calculated the energy spectrum having nonzero values only on the parameters of interest.
It is not possible to unequivocaly assign a physical mechanism to each term of the Hamiltonian because different mechanisms may contribute to the same term as explained above.
However, for simplicity, each term will be referred approximately according to its main effect.
Therefore, the Coulomb interaction between $f$-electrons ($\Fk{k}$) leads to splittings of the order of $10^4 \invcm$
and spin-orbit ($\spinZeta$) is typically $10^{3} \invcm$.
Configuration interaction ($\casimirAlpha$, $\casimirBeta$, and $\casimirGamma$) adds energy splittings of the order of $10^{2} \invcm$. 
The three-body terms can be very different among the ions, varying splittings within $(10^2\text{ - }10^3) \invcm$ from $3$ to $12$ electrons.
Also, the electrostatically-correlated-spin-orbit interaction ($\Pk{k}$) have smaller contributions of $\sim 10^2 \invcm$.
Magnetic contributions ($\Mk{k}$) are even smaller at about $10 \invcm$. 
Finally, the crystal field contribution is of the order of $10^2 \invcm$.
Table \ref{table:orders-of-mag} presents the range of values found for \LaF accross the lanthanide ions.

As mentioned, only energy differences are relevant, and it is customary to shift the spectrum to start at zero energy. 
However, a global shift is allowed by \bill and others for a better fitting. 
Here, this allowance is included as the parameter $\epsilon$, which other authors might include as an equivalent $\text{E}_\text{avg}$ parameter, e.g., \cheng.

\subsection{Orthogonal and non-orthogonal operators} 

In 1984, Judd, Crosswhite and Suskin \cite{Judd_Crosswhite_84,judd_complete_1984} discussed the usefulness of a complete set of orthogonal operators composing the parametric Hamiltonian.
The purpose of orthogonal operators is to remove correlations between the parameters during the fitting procedure, so that in the case of parameters included sequentially, the new and less important parameter will make minimal changes to the previous ones.
Consequently, the correlations when using non-orthogonal operators result in larger uncertainties in the parameters, as compared to the orthogonalized version \cite{newman_operator_1982}.
Judd \textit{et al.} derived the orthogonal operators related to the Coulomb interaction for both the single configuration $\hat{f}_k$ (with coefficients $F^{(k)}$) and the configuration interaction $\hat{e}_\alpha$, $\hat{e}_\beta$, $\hat{e}_\gamma$, and $\hat{t}_2$ (with coefficients, $\alpha$, $\beta$, $\gamma$, $T^{(2)}$, respectively).
In fact, the orthogonal operators for the single configuration description are better quantified in terms of the operators $\hat{e}_k^\perp$ (and coefficients $E_\perp^{(k)}$), as they obey simpler transformation properties with respect to $\casimir{\Gtwo} $ and $\casimir{\SO{7}}$ used to classify the states~\cite{wybourne1965spectroscopic}.
On the other hand, the same energy spectrum, for both the orthogonal and non-orthogonal version of the parametric Hamiltonian, can be obtained if one realizes a mapping between the set of coefficients related to each one.
As detailed in the Appendix \ref{app: orthogonalization}, this mapping can be written in the following form,
\begin{align}
    \OEk{1} &= E^{(1)} + \frac{4\alpha}{5}+\frac{\beta}{30}+\frac{\gamma}{25}, \label{eq: orthogonal E1}\\
    \OEk{2} &= E^{(2)}, \\
    \OEk{3} &= E^{(3)} -\frac{2\alpha}{5} +\frac{\Tk{2}(N-2)}{70\sqrt{2}},  \\
    \OcasimirAlpha &= \frac{4\alpha}{5}, \\
    \OcasimirBeta &= -4\alpha-\frac{\beta}{6}, \\
    \OcasimirGamma &= \frac{8\alpha}{5}+\frac{\beta}{15}+\frac{2\gamma}{25}, \\
    T^{(2)}_\perp & = T^{(2)} \label{eq: T2 prime},
\end{align}
with
\begin{align}
    E^{(1)} & =\frac{1}{9}\left(70F_{2}+231F_{4}+2002F_{6}\right), \label{eq: nonorthogonal E1}\\
    E^{(2)} & =\frac{1}{9}\left(F_{2}-3F_{4}+7F_{6}\right),\\
    E^{(3)} & =\frac{1}{3}\left(5F_{2}+6F_{4}-91F_{6}\right), \label{eq: nonorthogonal E3}
\end{align}
in which we used the subindex notation that relates to the superindex notation $F^{(k)} = D_k F_k$, as presented in Eq. (\ref{eqn:hamiltonian}), through the constant values $D_2 = 225$, $D_4 = 1089$, and $D_6 = 184041/25$. (See Table 6-2 in Ref. [\onlinecite{cowan1981theory}].)

We notice that, although the argument for using orthogonal operators was already established five years earlier, \bill decided to use the non-orthogonal operators in their calculations, as in Eq. (\ref{eqn:hamiltonian}).
However, later reference tables for three-body operators from Hansen, Judd, and Crosswhite \cite{HANSEN19961} provide matrix-element values for the orthogonalized operator $\left\langle \psi_{i}\right|\hat{t}_2^\perp\left|\psi_{j}\right\rangle$~\footnote{\qlanth includes the flexibility to use either orthogonal  or non-orthogonal operators.}, rather than for $\left\langle \psi_{i}\right|\hat{t}_2\left|\psi_{j}\right\rangle$.
Any attempt of using the $\Tk{2}$ parameters from Ref. [\onlinecite{carnall_systematic_1989}] as coefficient to the matrix elements in Ref. [\onlinecite{HANSEN19961}] will lead to wrong results.

At this point, we would like to highlight the special case of the $f^{12}$ configuration ($\text{Tm}^{3+}$) because it has been a point of misunderstanding and errors in past results.
Some matrix elements $\left\langle \psi_{i}\right|\op{t}_2\left|\psi_{j}\right\rangle_{f^{12}}$ of the non-orthogonal operator $\op{t}_2$ are nonzero; this is confusing since a three-body term is then present in $f^{12}$, which is a complementary configuration of two holes.
On the contrary, if one uses $\op{t}_2^\perp$ then all matrix elements $\left\langle \psi_{i}\right|\op{t}^\perp_2\left|\psi_{j}\right\rangle_{f^{12}}$ are zero, and, as such, the energy spectrum of $\text{Tm}^{3+}$ is independent of the parameter $T^{(2)}_\perp$.

We end this section mentioning that the transformation above is not complete, because the operators related to the $M$'s and $P$'s coefficients are not orthogonal between themselves. 
This lingering part of the Hamiltonian may be orthogonalized using the $\hat{z}_i$ operators initially described by Judd, Crosswhite, and Crosswhite in their description of the intra-atomic magnetic interactions \cite{judd_intra-atomic_1968} and explained in Ref. \cite{Judd-2008-JPB}. 
However, we will not consider it here, because it adds additional parameters that are not yet well explored in the literature. 
Thus, what we call the effective orthogonal Hamiltonian in this work is at best mostly orthogonal.

\section{Quantitative comparisons with the literature}
\label{sec: errors}

In this section, we will provide a quantitative comparison between results from \qlanth and the results presented by Carnall \etal for \LaF in Ref. [\onlinecite{carnall_systematic_1989}].
We will point to errors in the spectroscopic tables and additional features that allowed us to reproduce, fairly closely, their results.

\subsection{Errors in the spectroscopic tables of reduced matrix elements}

In 2008, Chen \etal~\cite{chen_few_2008} reported errors in the tabulated values of the reduced matrix elements of the spin-other-orbit coupling $\langle \hat{m}_k \rangle$ and the electrostatically correlated spin-orbit interaction $\langle \hat{p}_k \rangle$.
As Chen \etal described, errors were found in data files with tabulated reduced matrix elements, which we refer to as \fncross tables. 
These tables were implemented in the software SPECTRA (supported by the Argonne National Laboratory \cite{zhorinspectra}) and our results indicate that these tables were also used by \bill. 
The overall effects of these errors are responsible for differences of about $10 \invcm$.

In addition to the errors mentioned above, Judd and Lo  \cite{judd_three-electron_1994} identified errors in the reduced matrix elements of the three-electron configuration-interaction $\tk{k}$. 
We extracted the \textit{fncross} tables from Ref. \cite{linuxemp},
where there are two alternatives for the \fncross tables:  \textit{fn} and 
\textit{fn\_new}. 
The \textit{fn\_new} files fix the errors identified by Chen \etal, but still contain the errors identified by Judd and Lo. 
Furthermore, we found that the \textit{fncross} tables (\textit{fn} and \textit{fn\_new}) contain a few additional errors in the $\tk{k}$ matrix elements, listed in Table \ref{table:extra-errors-mini} (see Appendix \ref{sec: more-errors} for a complete table).
Finally, for $n=7$, there is a typo in Judd and Lo, where $\braopket{\LSterm{2}{\text{F5}}}{\tk{3}}{\LSterm{2}{\text{F8}}}$ should be $\braopket{\LSterm{2}{\text{F1}}}{\tk{3}}{\LSterm{2}{\text{F8}}}$.

Furthermore, when the method of fractional parentage is used to calculate reduced matrix elements for all $f^N$, some matrix elements that are exactly zero when computed with symbolic arithmetic (as done here) may yield small but non-zero values due to accumulated rounding error when using finite precision arithmetic.
Table \ref{table:extra-errors} in the appendix also includes notable errors due to the use of finite precision arithmetic in some matrix elements of $\hat{t}_k$.

\begin{center}
\tablefirsthead{
    \hline
    n & op & $\langle\text{LS}|$ & $|\text{LS'}\rangle$ & \qlanth & fncross \\
    \hline
    \noalign{\vskip 0.5ex}
}
\tablehead{
    \hline
    n & op & $\langle\text{LS}|$ & $|\text{LS'}\rangle$ & \qlanth & fncross \\
    \hline
    \noalign{\vskip 0.5ex}
}
\tabletail{
    \hline
    \multicolumn{6}{|c|}{continued $\cdots$} \\
    \hline
}
\tablelasttail{
    \hline
}
\bottomcaption{Additional errors  in the \textit{fncross} tables identified here. See Appendix \ref{sec: more-errors} for a complete table and color legends.
}

\small
\begin{supertabular}{|c|c|c|c|c|c|}
    \textcolor{red}{$3$} & \textcolor{red}{$\hat{t}_7$} & \textcolor{red}{${}^{2}{\text{L}}$} & \textcolor{red}{${}^{2}{\text{L}}$} & \textcolor{red}{-0.026503} & \textcolor{red}{-0.026053}  \\
    \noalign{\vskip 0.5ex} 
    \hline
    \noalign{\vskip 0.5ex} 
    \textcolor{red}{$6$} & \textcolor{red}{$\hat{t}_4$} & \textcolor{red}{${}^{1}{\text{S1}}$} & \textcolor{red}{${}^{1}{\text{S3}}$} & \textcolor{red}{\,\,5.737097} & \textcolor{red}{-5.737097} \\
    \textcolor{red}{$6$} & \textcolor{red}{$\hat{t}_4$} & \textcolor{red}{${}^{1}{\text{Q}}$} & \textcolor{red}{${}^{1}{\text{Q}}$} & \textcolor{red}{-0.856893} & \textcolor{red}{\,\,0.000000} \\
    \noalign{\vskip 0.5ex} 
    \hline
    \noalign{\vskip 0.5ex} 
    \textcolor{blue}{$7$} & \textcolor{blue}{$\hat{t}_3$} & \textcolor{blue}{${}^{2}{\text{F1}}$} & \textcolor{blue}{${}^{2}{\text{F8}}$} & \textcolor{blue}{-4.535574} & \textcolor{blue}{-3.239695} \\
    \noalign{\vskip 0.5ex} 
    \hline
    \noalign{\vskip 0.5ex} 
    \textcolor{red}{$8$} & 
    \textcolor{red}{$\hat{t}_3$} &
    \textcolor{red}{$\LSterm{1}{S1}$} &
    \textcolor{red}{$\LSterm{1}{S3}$} &
    \textcolor{red}{\,\,-5.7371} & \textcolor{red}{5.7371} \\
    \textcolor{red}{$8$} & 
    \textcolor{red}{$\hat{t}_6$} &
    \textcolor{red}{$\LSterm{1}{S2}$} &
    \textcolor{red}{$\LSterm{1}{S3}$} &
    \textcolor{red}{\,\,-3.55842} & \textcolor{red}{3.55842} \\
    \textcolor{red}{$8$} & 
    \textcolor{red}{$\hat{t}_7$} &
    \textcolor{red}{$\LSterm{1}{S3}$} &
    \textcolor{red}{$\LSterm{1}{S4}$} &
    \textcolor{red}{\,\,2.53546} & \textcolor{red}{-2.53546} \\
    \textcolor{red}{$8$} & 
    \textcolor{red}{$\hat{t}_2$} &
    \textcolor{red}{$\LSterm{1}{S2}$} &
    \textcolor{red}{$\LSterm{1}{S3}$} &
    \textcolor{red}{\,\,-0.589802} & \textcolor{red}{-1.4863} \\
    \textcolor{red}{$8$} & 
    \textcolor{red}{$\hat{t}_4$} &
    \textcolor{red}{$\LSterm{1}{Q}$} &
    \textcolor{red}{$\LSterm{1}{Q}$} &
    \textcolor{red}{\,\,0.856893} & \textcolor{red}{0.0} \\
    \textcolor{red}{$8$} & 
    \textcolor{red}{$\hat{t}_4$} &
    \textcolor{red}{$\LSterm{1}{S3}$} &
    \textcolor{red}{$\LSterm{1}{S4}$} &
    \textcolor{red}{\,\,-0.29277} & \textcolor{red}{0.29277} \\
    \noalign{\vskip 0.5ex} 
    \hline
    \noalign{\vskip 0.5ex} 
    \textcolor{red}{$12$} & \textcolor{red}{$\hat{t}_2$} & \textcolor{red}{${}^{1}{\text{G}}$} & \textcolor{red}{${}^{1}{\text{G}}$} & \textcolor{red}{\,\,-0.404061} & \textcolor{red}{\,\,0.000000} \\
    \hline
\end{supertabular}
\label{table:extra-errors-mini}
\end{center}

\begin{figure}
    \centering 
    \includegraphics[width=0.475\textwidth]{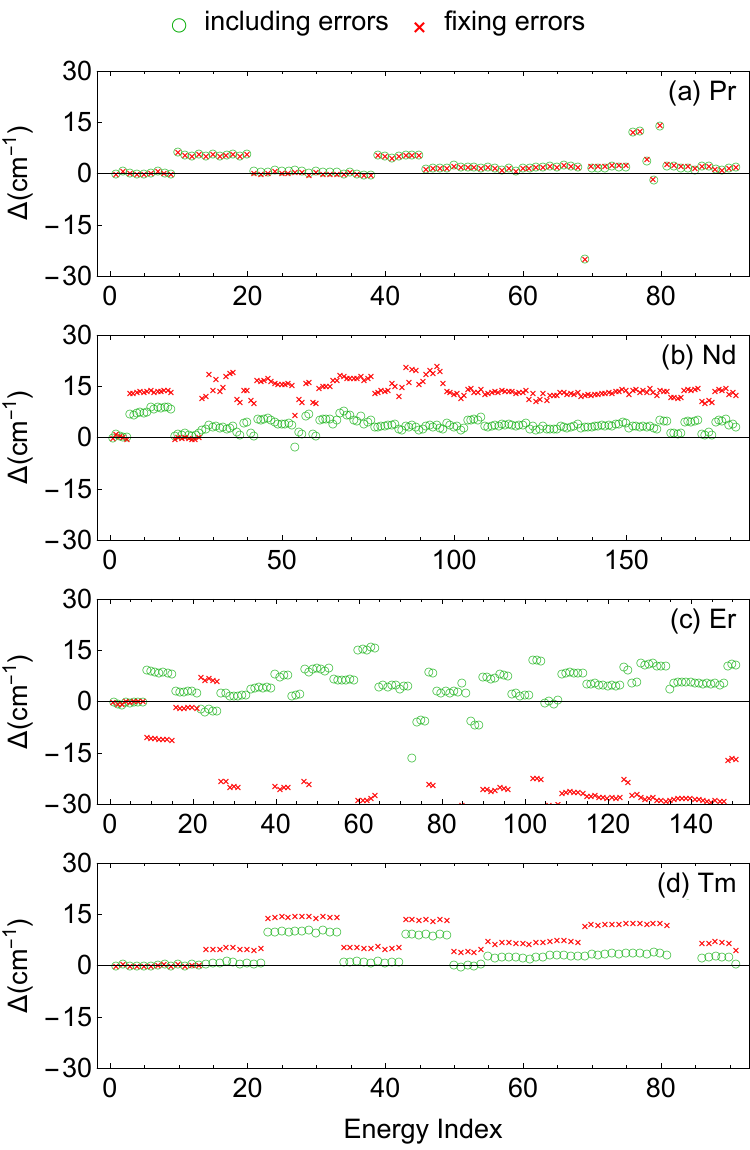}
    \caption{Differences ($\Delta \equiv  E_{ql} - E_R$) between the energies calculated using \qlanth ($E_{ql}$) and those quoted by Carnall \etal in Ref. [\onlinecite{carnall_systematic_1989}] ($E_R$) for the cases when the errors discussed in Section \ref{sec: errors} are included (green circle) and when they are not (red cross). 
   }
   \label{fig:calc_comparison}
\end{figure}

The results presented below makes us believe that the errors pointed above were present in the calculations performed by \bill. 
In Fig. \ref{fig:calc_comparison} we compare the calculated energies using \qlanth and those quoted by \carnall for the cases when the errors are included in the calculation and when they are not. 
We clearly see that the calculated energies are much closer to theirs when the errors are present.

\begin{figure}
    \centering
    \includegraphics[width=0.475\textwidth]{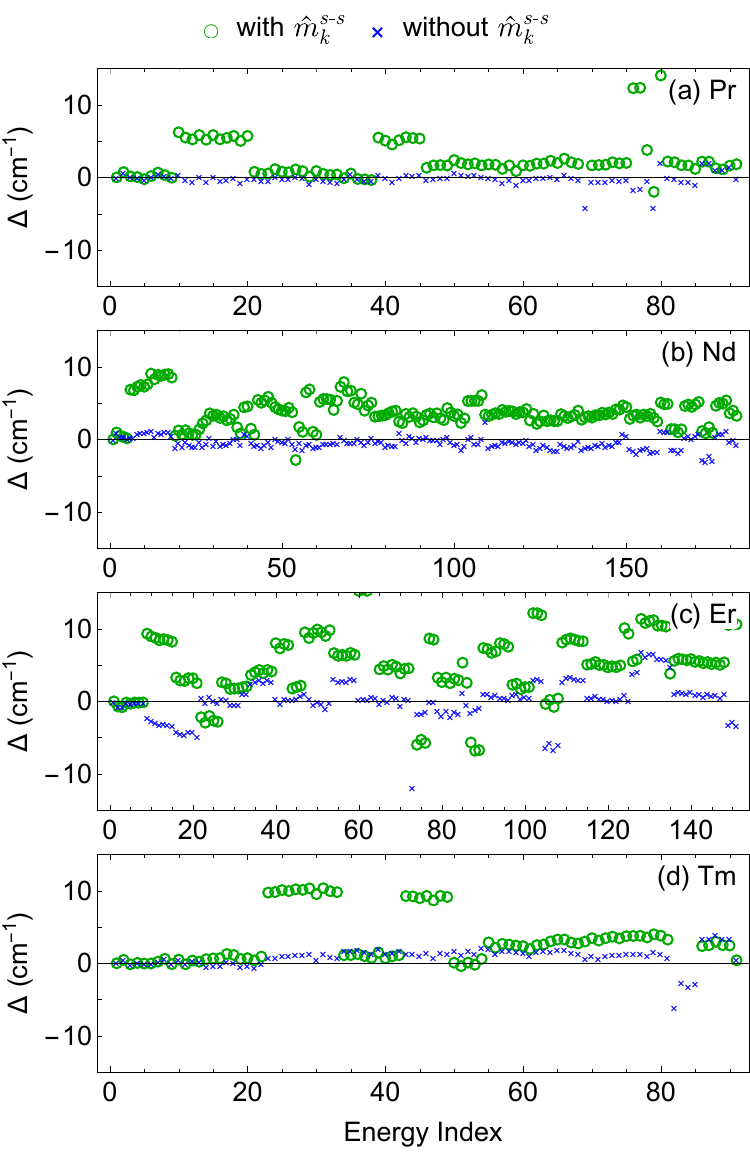}
       \caption{Differences ($\Delta \equiv  E_{ql} - E_R $) between the energies calculated using \qlanth ($E_{ql}$) and those quoted by Carnall \textit{et al.} in Ref. [\onlinecite{carnall_systematic_1989}] ($E_R$) for the cases when the Hamiltonian term $\hat{m}_{k}^{s\text{-}s}$ is included (green circle) and when it is excluded (blue cross).
       These results include the errors discussed in Section \ref{sec: errors}, thus the green circles are the same data as in  Fig. \ref{fig:calc_comparison}.
       }
\label{fig:qlanth_vs_Carnall_Pm_Dy_Tm}
\end{figure}

\subsection{Omission of the spin-spin interaction} 

Beyond the errors discussed above, we noticed that some of the remaining discrepancies in Fig. \ref{fig:calc_comparison} can be reduced by excluding the spin-spin contribution $\hat{m}_k^{s\text{-}s}$ to the Marvin integrals.
Whereas Carnall et al.~\cite{carnall_systematic_1989} claim to have included the spin-spin contribution to the magnetic interactions present in the spectrum, our calculations indicate that this term was omitted.
In Fig. \ref{fig:qlanth_vs_Carnall_Pm_Dy_Tm}, we compare the calculated energies using \qlanth and those quoted by Carnall \textit{et al.} for the cases when the spin-spin contribution is included in the calculation and when it is not.
Again, we clearly see that the calculated energies are much closer to reported values when the spin-spin contribution is absent. 
Notice that in this figure all the results include the errors discussed in the previous subsection.
Also, the effect of excluding the spin-spin contribution brings the differences down to $5\invcm$ or less.
Finally, we would like to stress that removing the spin-spin contribution was done with the intent of understanding the calculations in Ref. [\onlinecite{carnall_systematic_1989}]. 
However, we have no reason to believe that it should be removed, and therefore, except for the results in  Fig. \ref{fig:qlanth_vs_Carnall_Pm_Dy_Tm}, all other results presented here include this term.

\begin{figure*}
    \vspace{1cm}
    \begin{center}
    \includegraphics[width=0.95\textwidth]{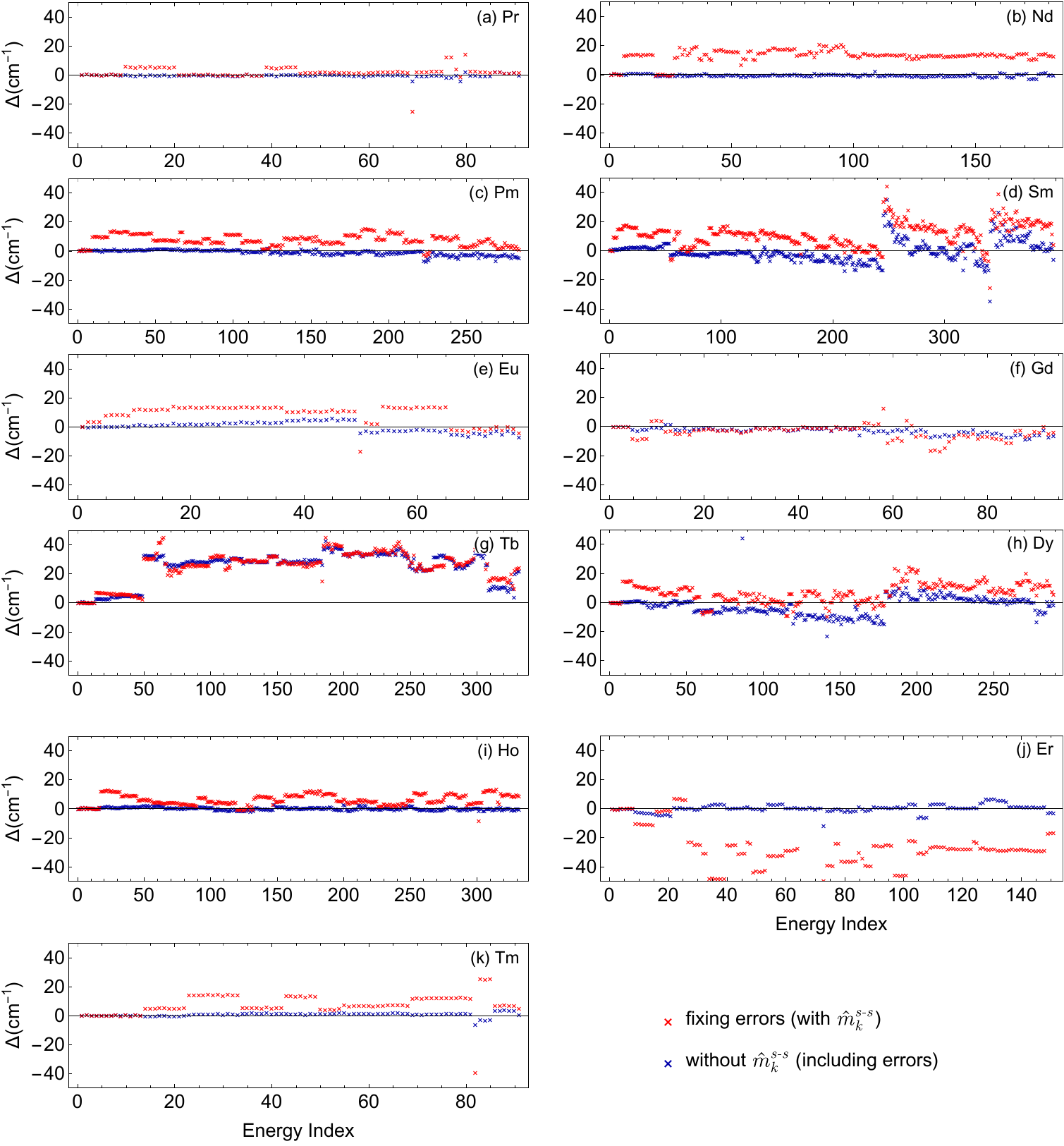}
    \caption{Differences ($\Delta \equiv  E_{ql} - E_R$) between the energies calculated using \qlanth ($E_{ql}$) and those quoted by Carnall et al. in Ref. [\onlinecite{carnall_systematic_1989}] ($E_R$) for the case when the errors discussed in Section \ref{sec: errors} are corrected and the Hamiltonian term $\hat{m}_{k}^{s\text{-}s}$ is included (red cross), and for the case when errors discussed in Section \ref{sec: errors} are included in the calculation and the Hamiltonian term $\hat{m}_{k}^{s\text{-}s}$ is excluded (blue cross). }
    \label{fig:qlanth_vs_Carnall_Ho_Dy_Er}
    \end{center}
\end{figure*}

A comparison between our best reproduction of \bill (blue cross in  Fig. \ref{fig:qlanth_vs_Carnall_Pm_Dy_Tm}) and calculations after fixing the errors (red cross in  Fig. \ref{fig:calc_comparison}) is presented in  Fig. \ref{fig:qlanth_vs_Carnall_Ho_Dy_Er}.
In this figure, most ions are reproduced to within a few $\!\invcm$, with $\trion{Tb}$ being the worst case, wherein starting at the fiftieth state there is a consistent discrepancy of  $\sim 30\invcm$, whose origin remains unexplained.

\subsection{Typographical mistakes in quoted energy levels}

In the appendix of Ref.  [\onlinecite{carnall_systematic_1989}], we have found a few obvious typographical errors. We detail them in Appendix \ref{sec: Carnall tables}. 
Furthermore, the tables sometimes report only the number of energy levels within a certain range, instead of listing the explicit values;  other times they report a gap in the spectrum at high energies. 
In a few cases, we found these ranges incompatible with calculations. 
For these reasons, we have included in the supplementary electronic files, the adjusted list of energies that we used to compare the results presented here.

\subsection{Other codes}
We performed an extensive analysis of our results with two other available codes. Specifically, we have made comparisons against \textit{Lanthanide} provided by Edvardsson and $\AA$berg \cite{edvardsson2001} and \linuxemp by Michael Reid \cite{linuxemp}.

The comparisons were established through the energy spectra, generated from different codes using the same set of parameters. 
First, it is important to note some evident differences between codes.
For instance, \textit{Lanthanide} does not include $\Mk{k}$ or $\Pk{k}$ parameters; therefore, we set those parameters to zero in these calculations. 
Overall, we found that \qlanth and \textit{Lanthanide} have an excellent numerical agreement (up to $10^{-7}$) only if the $\Tk{k}$ parameters are set to zero in the calculations.
On the other hand, calculations with a complete set of parameters, including $\Tk{k}$, show significant differences in the spectra.
Unfortunately, we could not resolve or find an intuitive explanation for these differences.
More details of these results are provided in the set of electronic files.

In the case of \linuxemp, we used a publicly available version from 2018 \cite{linuxemp}. 
This included two alternatives for the \fncross tables, including the latest one with fixes to the errors identified by Chen \etal \cite{chen_few_2008}. 
For comparison, we omitted the spin-spin contribution (which is also excluded in \linuxemp as per their documentation) and used the most recent version of the \fncross tables in \linuxemp, keeping the same parameters to calculate energies with both codes. 
In most cases the maximum absolute difference between the two sets of calculated energies (for the entire calculated energy spectra) was found to be smaller than $0.2\invcm$.
Exceptionally, in the cases of Eu, Gd, Tb, and Tm, the maximum absolute differences were $680 \invcm$, $32 \invcm$, $560 \invcm$, and  $94 \invcm$ respectively.
If the mean of the absolute differences is considered, then in all cases (excluding Tm, in which it is $16 \invcm$) the mean discrepancies are always smaller than $3\invcm$, which tells us that most of the energies were similar with a few exceptions.
We have identified that the outlier energies that give the large maximum discrepancies are due to the additional errors identified in Table \ref{table:extra-errors-mini}. 
Accounting for these errors, the maximum absolute differences drop to below $1.1\invcm$ in all cases.
Since \linuxemp is widely used, we provide corrected versions of the \fncross files in the supplementary materials.

\begin{center}
\begin{table*}
    \includegraphics[width=0.99\textwidth]{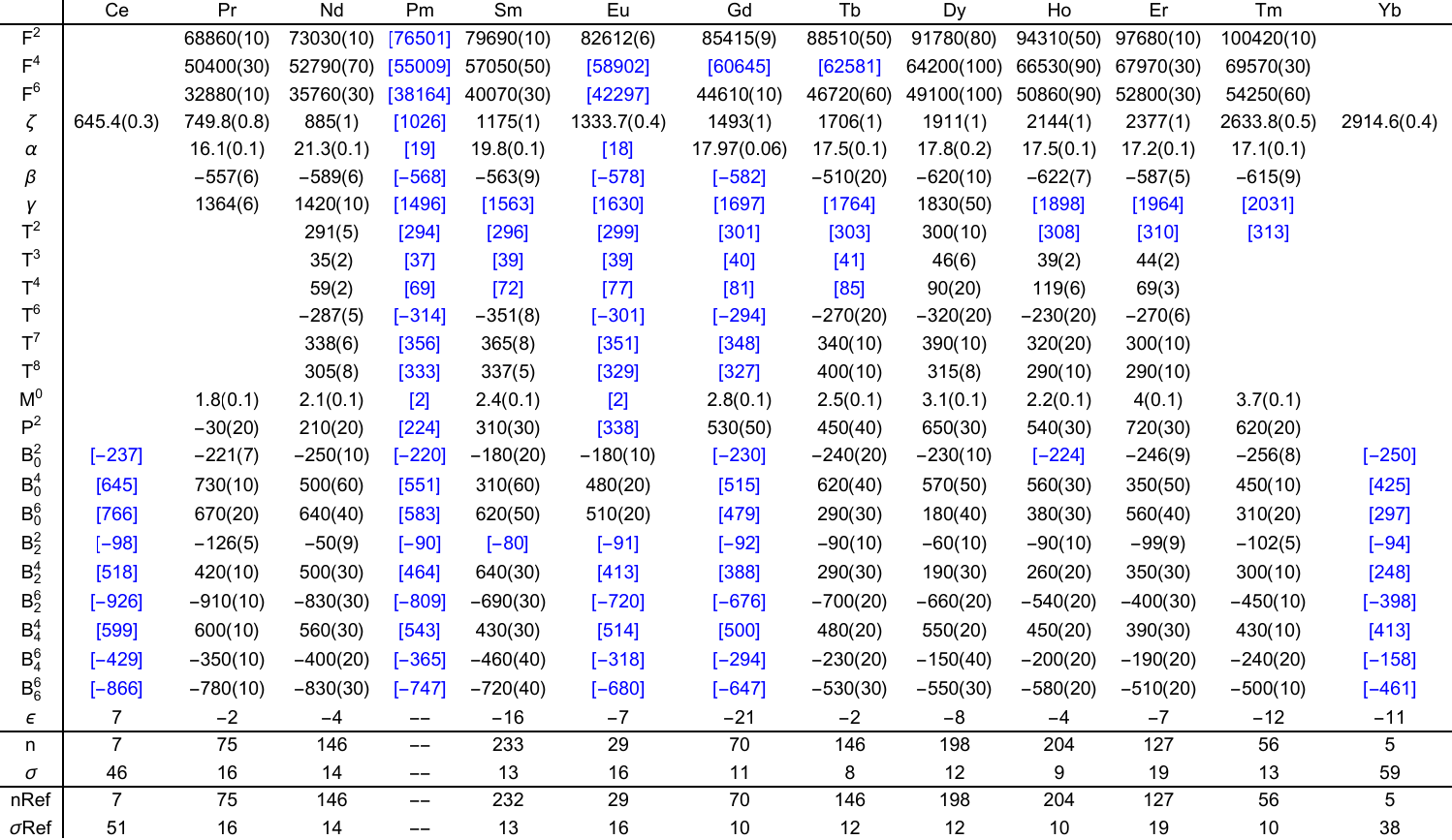}
    \caption{Fitted model parameters using \qlanth. Following \bill, parameters in brackets were held fixed during fitting. In the case of Eu, Gd, and Tb, some Slater parameters were constrained to be proportional to $\text{F}^2$. For Eu, $\text{F}^4 = 0.713 \text{F}^2$ and $\text{F}^6 = 0.512 \text{F}^2$. In the case of Gd, $\text{F}^4 = 0.71 \text{F}^2$. And in the case of Tb, $\text{F}^4 = 0.707 \text{F}^2$. The data was fitted in the following order: Pr, Nd, Dy, Ce, Sm, Ho, Er, Tm, Yb, Tb, Eu, Gd.}
       \label{table:qlanthLaF3}
\end{table*}
\end{center}

\section{Updated level structures}
\label{sec: updated level structure}

In this section we show our results of the re-fitted parameters of the \hamilton for the case of lanthanide ions in \LaF (fitted parameters for \liyorite are presented in Appendix \ref{appendix:liyorite}).
Specifically,  \tableref{table:qlanthLaF3} shows our obtained parameters when the model is fitted to the same energy levels used in Carnall et al. [\onlinecite{carnall_systematic_1989}]. In these calculations, we have included the spin-spin contribution term $\hat{m}_{k}^{s\text{-}s}$ and corrected the errors discussed in Section \ref{sec: errors}. 
The assumptions for these calculations are described below in detail and reflect our best effort to reproduce the parameter choices in that reference.

\subsection{Fitting assumptions}

First, our fittings to the model parameters use the same experimental energy levels used in Ref. \cite{carnall_systematic_1989} for \LaF.
Second, previous numerical approaches used a truncated version of the \hamilton in the fitting procedure (although the full diagonalization is made with the converged parameters to obtain the final energy spectrum); here truncation was completely avoided.
Third, when some parameters are constrained as proportional to others, we kept the same ratios. 
For instance, for Eu we fixed $\Fk{4} = 0.713 \Fk{2}$ and $\Fk{6} = 0.512 \Fk{2}$; for Gd we fixed $\Fk{4} = 0.71 \Fk{2}$; and for Tb we fixed $\Fk{4} = 0.707 \Fk{2}$.
For the $\Mk{k}$ and $\Pk{k}$ parameters we used the constraints $\Mk{2} = 0.56 \Mk{0}$, $\Mk{4} = 0.31 \Mk{0}$, $\Pk{4} = 0.5 \Pk{2}$, and $\Pk{6} = 0.1 \Pk{2}$ for all ions. 
Fourth, since \carnall held some parameters fixed, we repeated this pattern (see Table \ref{table:qlanthLaF3}).

In \bill, ad-hoc choices were made for some the parameters that were held fixed with the intent of improving the fitting. 
Instead of addressing these small improvements individually, we chose to implement a systematic protocol as explained in the following.
When a parameter is held fixed for a given lanthanide, previously fitted values in other lanthanides are used to decide the value at which it is fixed during fitting.
We notice that for Pr, Nd, and Dy every parameter is freely varied, so we used those three cases as starting points.
From there, we used a linear fit (number of electrons in the $f$-shell versus parameter value) to fix the parameter values for Ce and Sm. 
For example, a linear fit from the $\Bkq{2}{2}$ parameters from Pr, Nd and Dy [$(2,-126)$, $(3,-50)$, and $(9,-60)$] allowed us to extrapolate/interpolate and obtain the values for Ce and Sm [$(1,-98)$ and $(5,-80)$].
We followed this procedure in  the following sequence: Ce $\rightarrow$ Sm $\rightarrow$ Ho $\rightarrow$ Er $\rightarrow$ Tm $\rightarrow$ Yb $\rightarrow$ Tb $\rightarrow$ Eu $\rightarrow$ Gd. 
In this sequence, whenever a parameter is to be held fixed for a given lanthanide, the previously fitted values are considered.
For example, the $\Bkq{2}{0} = -224 \invcm$ parameter in Ho was obtained from a linear fit from Pr, Nd, Dy, and Sm (notice that we have not used $\Bkq{2}{0}$ from Ce because in that case $\Bkq{2}{0}$ was fixed and not freely varied).
The $\Fk{k}$ and $\spinZeta$ parameters are exceptions to the linear fit, these parameters have a non-linear trend across the lanthanide row.
For these parameters we took an empirical approach and used a cubic spline interpolation based on previously fit values.
A common practice is using $\spinZeta \propto Z^2$ \cite{khomskii_revieworbital_2022}, where $Z$ is the atomic number.
With this protocol, we completed the table in the following order: Ce, Sm, Ho, Er, Tm, Yb, Tb, Eu, and Gd.
This order was motivated by the number of estimated parameters needed and number of experimental values available.

The above paragraph describes our best judgment in reproducing and comparing results with \bill. 
However, we acknowledge that different approaches could further improve the fitting process. 
For instance, we believe the parameters follow a trend across the lanthanide series, making a global fit -- fitting the parameters for all lanthanides simultaneously -- a more suitable approach. 
This could involve imposing a simple relationship between the parameters and reducing their overall number of independent variables. 
As discussed earlier, it is important to perform the fitting in stages, starting with the largest contributions and progressively incorporating corrective terms in subsequent steps.

\begin{figure}
    \centering 
    \includegraphics[width=0.475\textwidth]{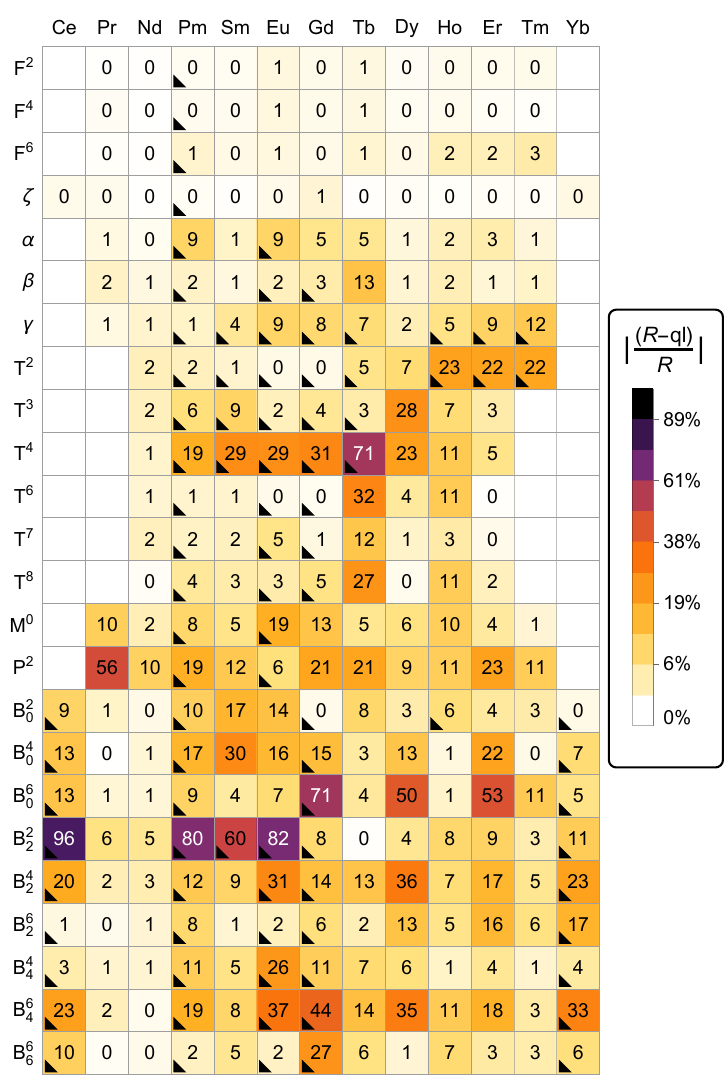}
    \caption{Relative absolute differences (\%) between the fitted parameters shown in \tableref{table:qlanthLaF3} (ql) and those quoted by Carnall et al. in Ref. [\onlinecite{carnall_systematic_1989}] (R). The colors represent the intensity of these percentage differences according to the scale bar on the right and, for better visualization, we also write the calculated values on their corresponding squares. Empty squares, not numbered, indicate parameters that are not present in the model.}
   \label{fig:param_ratios_delta_cl}
\end{figure}

\begin{figure}
   \centering
   \includegraphics[width=0.9\columnwidth]{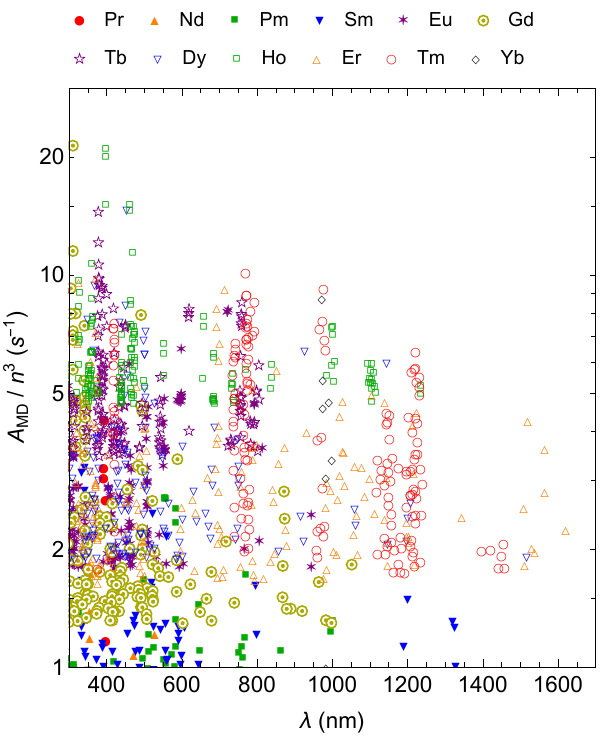}
   \caption{Calculated magnetic dipole transition rates $A_\text{MD}/n^3$ of the lanthanide ions in \LaF for  transitions from ultraviolet to infra-red. All data provided in the electronic files.}
   \label{fig:AMD_rates}
\end{figure}

\subsection{Fitting method and uncertainties}

Reproducing the approach of \bill, we used the Levenberg-Marquard method \cite{levenberg1944method,marquardt1963algorithm} to find parameter values that minimize $s^2 = \sum_i \left( E_{\text{ql}} - E_{\text{EXP}} \right)_i^2$. 
This is a deterministic fitting method, susceptible to the intial guess, so we use the parameters in Ref. \cite{carnall_systematic_1989} as starting points.
As such, the fitted parameters here reflect a refinement of those results, rather than an attempt to find global minima, see discussion in Appendix \ref{app: global fitting}.

For obtaining the uncertainties, the specific parametric form of the \hamilton allows one to compute the gradient of each energy $i$ as a function of the parameters $E_i ( \{ \boldsymbol{p} \}) = E_{i}^0 + \nabla E_i \cdot (\boldsymbol{p}_0  - \boldsymbol{p})$, where $\boldsymbol{p}$ comprises the set of parameters involved in the fitting. 
This method requires only a simple re-interpretation of terms in first-order perturbation theory.
This linear approximation of the eigenvalues around the local minimum then allows one to compute $s^2$ to second order in the model parameters $\boldsymbol{p}$. 
Here, a constant uncertainty in the energy differences was assumed to be $1\invcm$.
This is compatible with the nominal uncertainty of the energies in Ref. \cite{carnall_systematic_1989}, and provides uncertainties in the parameters that are of similar magnitude to the ones provided by them.
Together with the number of degrees of freedom $\nu$ in the corresponding fit, this allows us to compute the error intervals from the contour  $\tilde{\chi}^2=\frac{1}{\nu}\frac{s^2}{\sigma_\text{exp}^2}= \tilde{\chi}^2_\text{min}+1$ \cite{bevington2003data}.

\subsection{Results}

Our main results are summarized in  \tableref{table:qlanthLaF3}, which shows fitted parameters for \LaF (see Appendix \ref{appendix:liyorite} for fitted parameters in \liyorite).
Analysing these results, we find that the average root mean squared deviation ($\sigma$) in our calculation is similar to those ($\sigma_\text{Ref}$) in Ref. [\onlinecite{carnall_systematic_1989}], despite the significant differences in some parameters.
Fig. \ref{fig:param_ratios_delta_cl} shows the percentage differences between fitted parameters. 
Investigating the results in this figure, we notice that the electrostatic interaction $\Fk{2}$, $\Fk{4}$, $\Fk{6}$, and the spin-orbit $\spinZeta$ parameters do not show a significant change with respect to the reference values. 
The two-electron configuration-interaction parameters $\casimirAlpha$, $\casimirBeta$, and $\casimirGamma$ show small relative differences ($\lesssim 10\%$). 
This is understandable because these parameters (for which there are no significant errors in \bill) set  the general energy scale. 
Larger differences (about $30\%$ or higher) appear in the three-electron configuration-interaction $\Tk{k}$ and the pseudo-magnetic $\Pk{k}$ parameters.
Interestingly, $\Tk{2}$ parameters become significantly different above $N=7$ electrons and the $\Tk{4}$ parameter have the largest relative differences, across the lanthanide row.
The crystal field parameters $\Bkq{k}{q}$ are generically different, with the exception of Pr and Nd ions that show surprisingly good agreement.
In fact, all Nd fitted parameters are very close to the reference results.

\begin{center}
   \begin{figure*}
       \includegraphics[width=0.95\textwidth]{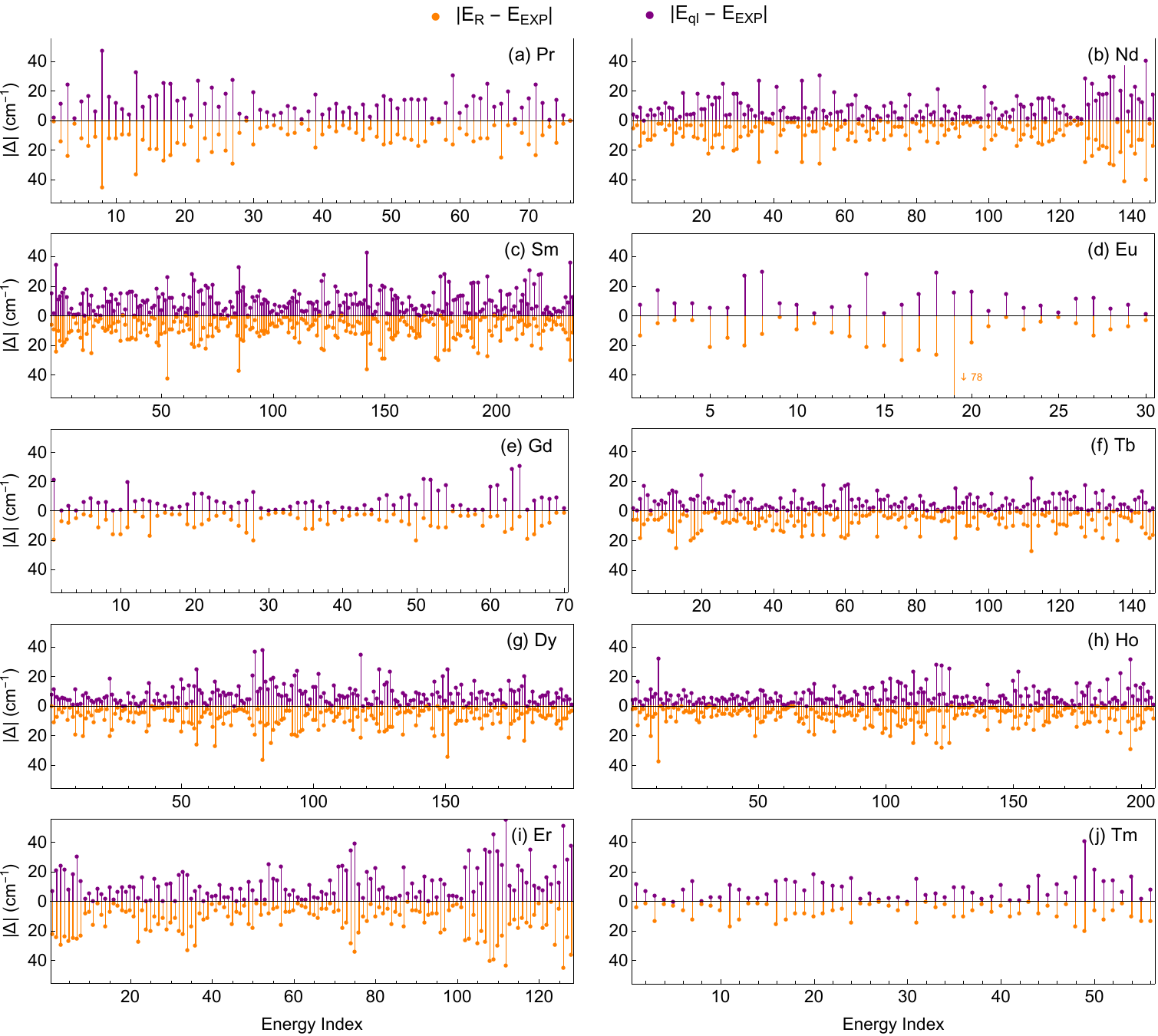}
       \caption{Absolute differences between calculated values with refitted parameters and experimental data. The upper axis (purple) shows the difference between the our calculated values and the experimental ones. The lower axis (orange), shows the differences between the values calculated by \carnall and the same experimental ones.}
       \label{fig:exp_comparison}
   \end{figure*}
\end{center} 

For completenes, Fig. \ref{fig:exp_comparison} compares the difference between the calculated energies and experimental values in ~\bill with our newly fit results that includes all the corrections noted above.
Note that the residuals between the fitted energies and the experimental values in  Fig. \ref{fig:exp_comparison} are $\sim 10 \invcm$.
Quantitatively, this agrees with the root mean square error ($\sigma$) of $8\invcm \text{ to } 19\invcm$ shown in Table \ref{table:qlanthLaF3}, except in the cases of Ce and Yb which have exceptionally large $\sigma$ due to the large number of parameters in comparison with the number of available energy levels. 
Relatedly, we notice that the Pr, Eu, Gd, and Tm ions present less than $100$ energies for the fitting; in the case of Eu there are only $\sim 30$, which can be the reason for the large differences in the resulting parameters, as shown in Fig. \ref{fig:param_ratios_delta_cl}.

\section{Magnetic dipole transitions}
\label{sec: magnetic dipole transitions}

In this section we discuss the absorption and emission rates for  magnetic dipole transitions \footnote{\qlanth also implements calculations of forced electric dipole transitions.}, and show results for the lanthanides in \LaF. 
These calculations expand upon and correct the work of Dodson and Zia in Ref.~[\onlinecite{dodson_2012}] that omitted the $\Tk{2}$ contribution for $\trion{Tb}$ in $4f^{12}$ and also contained a typo related to $\casimirAlpha$ that contributed to non-diagonal matrix elements to $\hat{L}^2$.
Both absorption and emission are calculated from the line strength defined between any two energy levels as
\begin{align}
    \hat{\cal S} \left( \phi_i,\phi_f \right) = \left| \left< \phi_i \right| \hat{\boldsymbol \mu}\left| \phi_f \right> \right|^2 ,
\end{align}
where $\hat{\boldsymbol \mu} = -\mu_B \left( \hat{\boldsymbol L}+ g_s \hat{\boldsymbol S} \right)$ is the magnetic dipole operator and $\phi_i,\phi_f$ are the corresponding wavevector of the energy levels $E_i,E_{f}$.
The spontaneous emission rate is given (in units of $s^{-1}$) by 
\begin{equation}
    \frac{A_\text{MD}\!\left( \phi_i,\phi_f \right) }{n^3}  = \frac{16 \pi^3 \mu_0  \hat{\cal S}\!\left( \phi_i,\phi_f \right)} {3h \lambda^3 g_i},\quad \text{with } E_i > E_f
\end{equation}
where $\lambda = hc / |E_i - E_f|$ is the vacuum-equivalent wavelength of the transition, $g_i$ is the degeneracy of the highest energy level, and $n$ is the refraction index of the host material.
The oscillator strength is computed as
\begin{align}
    \frac{f_\text{MD} \left( \phi_i,\phi_f \right) }{n}  = \frac{8 \pi^2 m_e  \hat{\cal S}\left( \phi_i,\phi_f \right)} {3hc e^2 \lambda g_i},\quad \text{with } E_i < E_f,
\end{align}
where $g_i$ is the degeneracy of the lowest energy level. The $\mu_0$, $h$, $m_e$, $c$, and $e$ are fundamental physical constants.

Figure \ref{fig:AMD_rates} shows the ultraviolet to near-infrared spontaneous emission rates above 1/s for all lanthanides in \LaF. 
The energies involved in each transition are not shown; however, the complete information is included in the attached electronic files.
It is interesting to notice that the largest transition rates happen for $N \geq 7$ (hollow symbols), thus from Gd to Yb in the lanthanide row.
These results are comparable to previous free-ion calculations of MD emission rates in the lanthanides \cite{dodson_2012}, but here the rates are calculated for transitions between pairs of individual crystal-field states.

\begin{center}
   \begin{figure*}
       \includegraphics[width=0.99\textwidth]{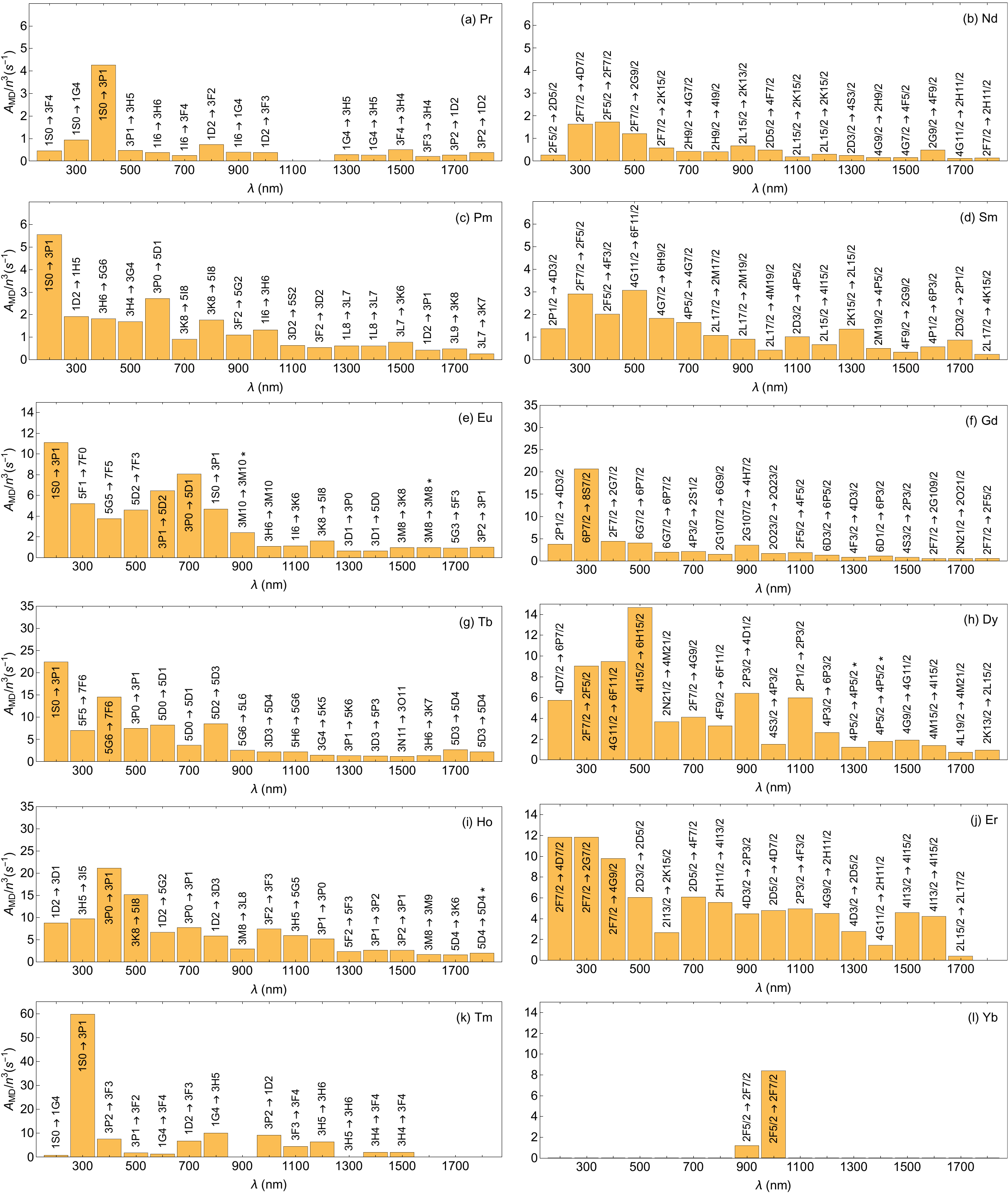}
       \caption{Maximum magnetic dipole transition rates for different wavelength ranges across the lanthanide ions. The height of the bar indicates the magnitude of the rate and the text shows the main $(2\text{S}+1)\text{LJ}$ component of the initial and final states. Cases marked with ${}^*$ are cases in which the given coarse labels are ambiguous, see Table \ref{tbl: eigenstates} for a longer description of these cases.}
       \label{fig:max-AMD-rates}
   \end{figure*}
\end{center}

In Fig. \ref{fig:max-AMD-rates}, we provide more details of some transitions by showing only the largest transition rate within wavelength intervals.
In this figure the bars represent the transition rate and carry coarse labels representing the states involved in that transition.
The coarse label shows only the largest contribution to the eigenvector, using $(2S+1)LJ$ notation and the $m_J$ is omitted.
Notice that the left column contains the lanthanide ions that have integer spins, while the right column shows the ions with half-integer spins.
Of particular interest is the case of Er around $1540 \text{ nm }$ transition.
Our calculations show that the important transition $\LSJterm{4}{I}{15/2} \rightarrow \LSJterm{4}{I}{13/2}$ at the telecommunication band, with $\lambda = 1543.28 \textrm{ nm}$, has $A_\text{MD}/n^3 = 4.56 \text{ s}^{-1}$ (comparable to erbium-doped \LaF transparent gel \cite{10.1063/1.1629772}).

Table \ref{tbl: eigenstates} shows in detail the energies and eigenvectors involved in some of the spontaneous emissions shown above.
Specifically, we show those cases in which the transitions occur between energy states that differ only by their $m_J$ component.

\section{Conclusions}

This work reviews in detail the calculation of the energy spectra and wavefunctions of trivalent lanthanide ions in crystal hosts using the \hamilton.
We also review the theoretical path that transforms the Hamiltonian from the notation of solid-state physics to the form finally used for calculations.
This form is limited to the ground configuration of $4f$-electrons, although it also includes configuration-interaction corrections.
We integrate into our analysis data gathered from previous literature \cite{carnall_systematic_1989, chen_few_2008, judd_three-electron_1994, judd_intra-atomic_1968,velkov_multi-electron_2000,judd_complete_1984}.
In addition, by reviewing the details of using orthogonal operators, we have resurfaced a worthwhile discussion of how to best parametrize the \hamilton.

We detail the computational challenges encountered in replicating a canonical paper on lanthanide spectroscopy and provide an open source code, \qlanth, to ensure reproducibility.
This code offers the Hamiltonian representation in both orthogonal and non-orthogonal parameters. 
It can calculate line strengths and transition rates due to magnetic dipole and, going beyond the scope of this work, also implements the Judd-Ofelt approach for forced electric dipole transitions. 
Additionally, it has the ability to include or exclude the spin-spin contribution, and enable or disable the errors in spectroscopic tables~\cite{chen_few_2008}.

Updated calculations were compared against those from \bill, clarifying some of their implicit assumptions and finding some additional errors in widely used spectroscopic tables.
In particular, our results suggest that the spin-spin dipolar interaction between $f$ electrons was omitted from their calculations, despite \carnall's claim of having included it.
We also compared our results against other codes \cite{linuxemp,edvardsson2001}.
Lastly, we provided an updated version of the description of lanthanide ions in \LaF, as well as providing exhaustive data describing magnetic dipole transitions across the lanthanide row.

In the past decade, experimental advancements have enabled increasingly detailed observation and manipulation of lanthanide ions in crystals, from the coherent control and optical readout of single ions \cite{siyushev_coherent_2014} to addressing nuclear spins with coherence times extending into hours \cite{zhong_optically_2015}, as well as refined descriptions of their hyperfine structure \cite{jobbitt_prediction_2021}.
Given that the \hamilton plays a central role in the theoretical description of these phenomena and considering that necessary computational details are required for its actual usage, this work offers a computational foothold for future research involving lanthanide ions.

\section*{Acknowledgments}

We are grateful to Christopher M. Dodson for useful discussions and sharing iterations of his prior calculations, as well as to Lee Basssett for his insightful feedback. This paper and the development of \qlanth were supported by the National Science Foundation through grants DMR-1921877 and DMR-1922025.

\renewcommand{\arraystretch}{1.7}
\begin{table*}[t]
\begin{tabular}{|c|c|c|}
\hline 
$\begin{array}{c}
    \text{Lanthanide}\\
    \lambda (\text{nm})
\end{array}$ & 
$\begin{array}{c}
    E_\text{initial} (\invcm)\\
    E_\text{final} (\invcm)
\end{array}$ &
$\begin{array}{c}
    \left|\psi_\text{initial}\right\rangle \\
    \left|\psi_\text{final}\right\rangle
\end{array}$
\tabularnewline
\hline 
\hline 
$\begin{array}{c}
    \text{Eu}\\
    948\text{ nm}
\end{array}$ & 
$\begin{array}{c}
    55717\\
    45168
\end{array}$ & 
$\begin{array}{c}
    -0.23\ket{{}^{3}{\text{M2}_{10,\pm 6}}}+0.21\ket{{}^{3}{\text{M2}_{10,0}}}-0.20\ket{{}^{3}{\text{M3}_{10,\pm 6}}}-0.20\ket{{}^{1}{\text{N2}_{10,\pm 6}}} \\
    +0.22\ket{{}^{3}{\text{M3}_{10,\pm 5}}}-0.21\ket{{}^{3}{\text{O}_{10,\pm 5}}}
\end{array}$\tabularnewline
\hline 
$\begin{array}{c}
\text{Eu}\\
1568\text{ nm}
\end{array}$ & $\begin{array}{c}
80943\\
74565
\end{array}$ & $\begin{array}{c}
-0.22\ket{{}^{3}{\text{M1}_{8,\pm 8}}}+0.21\ket{{}^{3}{\text{M1}_{8,\pm 2}}} \\
-0.23\ket{{}^{3}{\text{M1}_{8,\pm 7}}}+0.17\ket{{}^{1}{\text{L4}_{8,\pm 7}}}
\end{array}$\tabularnewline
\hline 
$\begin{array}{c}
\text{Dy}\\
1339\text{ nm}
\end{array}$ & $\begin{array}{c}
46383\\
38915
\end{array}$ & $\begin{array}{c}
-0.23\ket{{}^{4}{\text{P2}_{\frac{5}{2},\frac{5}{2}}}}-0.21\ket{{}^{2}{\text{F7}_{\frac{5}{2},\frac{5}{2}}}}-0.17\ket{{}^{4}{\text{P2}_{\frac{5}{2},-\frac{3}{2}}}}-0.17\ket{{}^{4}{\text{G4}_{\frac{5}{2},\frac{5}{2}}}} \\
-0.4\ket{{}^{4}{\text{P2}_{\frac{5}{2},-\frac{3}{2}}}}-0.32\ket{{}^{6}{\text{P}_{\frac{5}{2},-\frac{3}{2}}}}+0.28\ket{{}^{4}{\text{G1}_{\frac{5}{2},-\frac{3}{2}}}}+0.28\ket{{}^{4}{\text{G4}_{\frac{5}{2},-\frac{3}{2}}}}
\end{array}$\tabularnewline
\hline 
$\begin{array}{c}
\text{Dy}\\
1352\text{ nm}
\end{array}$ & $\begin{array}{c}
46477\\
39078
\end{array}$ & $\begin{array}{c}
+0.21\ket{{}^{4}{\text{P2}_{\frac{5}{2},\frac{3}{2}}}}-0.21\ket{{}^{4}{\text{P2}_{\frac{5}{2},-\frac{5}{2}}}}+0.20\ket{{}^{4}{\text{I2}_{\frac{9}{2},-\frac{1}{2}}}}+0.19\ket{{}^{2}{\text{F7}_{\frac{5}{2},\frac{3}{2}}}}-0.19\ket{{}^{2}{\text{F7}_{\frac{5}{2},-\frac{{5}}{2}}}} \\
+0.39\ket{{}^{4}{\text{P2}_{\frac{5}{2},-\frac{5}{2}}}}+0.31\ket{{}^{6}{\text{P}_{\frac{5}{2},-\frac{5}{2}}}}-0.28\ket{{}^{4}{\text{G1}_{\frac{5}{2},-\frac{5}{2}}}}-0.28\ket{{}^{4}{\text{G4}_{\frac{5}{2},-\frac{5}{2}}}}
\end{array}$\tabularnewline
\hline 

$\begin{array}{c}
\text{Ho}\\
1792\text{ nm} 
\end{array}$ & $\begin{array}{c}
41651\\
36071
\end{array}$ & $\begin{array}{c}
+0.43\ket{{}^{5}{\text{D}_{4,0}}}-0.36\ket{{}^{3}{\text{F2}_{4,0}}}+0.31\ket{{}^{3}{\text{F4}_{4,0}}}-0.27\ket{{}^{5}{\text{D}_{4,\pm 2}}} \\
+0.33\ket{{}^{5}{\text{D}_{4,1}}}-0.33\ket{{}^{5}{\text{D}_{4,-1}}}+0.24\ket{{}^{3}{\text{H4}_{4,1}}}-0.24\ket{{}^{3}{\text{H4}_{4,-1}}} 
\end{array}$ \tabularnewline 
\hline 
\end{tabular}
\caption{Energies and wavevectors truncated to the largest contributions involved in the spontaneous transitions shown in Fig. \ref{fig:max-AMD-rates} marked with ${}^*$. For each transition, the transition wavelength $\lambda = ( E_\text{initial} - E_\text{final} )^{-1}$ is provided together with the initial and final energies, $E_\text{initial}$ and $E_\text{final}$, as well as the corresponding eigenvectors $\left| \psi_\text{initial} \right >$ and $\left| \psi_\text{final} \right >$. Contributions to the eigenvectors different only by the sign of $m_J$ are aggregated as $\pm m_J$.}
\label{tbl: eigenstates}
\end{table*}

\newpage
\bibliographystyle{apsrev4-2}
\bibliography{refined_refs}

\appendix

\section{Additional errors in the \fncross tables}
\label{sec: more-errors}

Three versions of the fncross files are provided in the supplemenatary materials. Two versions (\textit{fn} and \textit{fn\_new}) are identical to those included in the 2018 version of \linuxemp \cite{linuxemp}, and the third version \textit{fn\_ql} is one in the same format that includes all the corrections described in this paper. 

\begin{center}
\tablefirsthead{
    \hline
    n & op & $\langle\text{LS}|$ & $|\text{LS'}\rangle$ & \qlanth & fncross \\
    \hline
    \noalign{\vskip 0.5ex}
}
\tablehead{
    \hline
    n & op & $\langle\text{LS}|$ & $|\text{LS'}\rangle$ & \qlanth & fncross \\
    \hline
    \noalign{\vskip 0.5ex}
}
\tabletail{
    \hline
    \multicolumn{6}{|c|}{continued $\cdots$} \\
    \hline
}
\tablelasttail{
    \hline
}
\bottomcaption{Errors in $\hat{f}_k$ and $\tk{k}$ in the \textit{fncross} tables. The values that given as $0.000000$ in the \qlanth column, are exactly 0. The row in blue was incorrectly attributed to $\langle {}^{2}{\text{F3}} |\tk{3}| {}^{2}{\text{F8}} \rangle$ by Judd and Lo \cite{judd_three-electron_1994} and the rows in red were not quoted by them.}

\small
\begin{supertabular}{|c|c|c|c|c|c|}
    \textcolor{red}{$3$} & 
    \textcolor{red}{$\hat{t}_7$} & 
    \textcolor{red}{${}^{2}{\text{L}}$} & 
    \textcolor{red}{${}^{2}{\text{L}}$} & 
    \textcolor{red}{-0.026503} & 
    \textcolor{red}{-0.026053} \\
    \noalign{\vskip 0.5ex} 
    \hline
    \noalign{\vskip 0.5ex} 
    $6$ & $\hat{f}_2$ & ${}^{3}{\text{I1}}$ & ${}^{3}{\text{I5}}$ & \,\,0.000000 & -0.000001 \\
    $6$ &  & ${}^{3}{\text{I1}}$ & ${}^{3}{\text{I6}}$ & \,\,0.000000 & -0.000002 \\
    $6$ &  & ${}^{3}{\text{I2}}$ & ${}^{3}{\text{I6}}$ & \,\,0.000000 & \,\,0.000001 \\
    $6$ &  & ${}^{3}{\text{I3}}$ & ${}^{3}{\text{I6}}$ & \,\,0.000000 & -0.000001 \\
    $6$ &  & ${}^{3}{\text{K2}}$ & ${}^{3}{\text{K6}}$ & \,\,0.000000 & -0.000002 \\
    $6$ & $\hat{f}_4$ & ${}^{3}{\text{I2}}$ & ${}^{3}{\text{I6}}$ & \,\,0.000000 & \,\,0.000001 \\
    $6$ &  & ${}^{3}{\text{I3}}$ & ${}^{3}{\text{I6}}$ & \,\,0.000000 & \,\,0.000001 \\
    $6$ & $\hat{f}_6$ & ${}^{3}{\text{I1}}$ & ${}^{3}{\text{I6}}$ & \,\,0.000000 & \,\,0.000001 \\
    $6$ &  & ${}^{3}{\text{I2}}$ & ${}^{3}{\text{I6}}$ & \,\,0.000000 & \,\,0.000002 \\
    $6$ &  & ${}^{3}{\text{I3}}$ & ${}^{3}{\text{I6}}$ & \,\,0.000000 & -0.000001 \\
    $6$ &  & ${}^{3}{\text{K2}}$ & ${}^{3}{\text{K6}}$ & \,\,0.000000 & \,\,0.000001 \\
    $6$ &  & ${}^{3}{\text{K4}}$ & ${}^{3}{\text{K4}}$ & -0.000106 & -0.000104 \\
    $6$ & $\hat{t}_2$ & ${}^{1}{\text{S2}}$ & ${}^{1}{\text{S3}}$ & -0.448250 & \,\,0.448249 \\
    \textcolor{red}{$6$} & \textcolor{red}{$\hat{t}_3$} & \textcolor{red}{${}^{1}{\text{S1}}$} & \textcolor{red}{${}^{1}{\text{S3}}$} & \textcolor{red}{\,\,5.737097} & \textcolor{red}{-5.737097} \\
    \textcolor{red}{$6$} & \textcolor{red}{$\hat{t}_4$} & \textcolor{red}{${}^{1}{\text{Q}}$} & \textcolor{red}{${}^{1}{\text{Q}}$} & \textcolor{red}{-0.856893} & \textcolor{red}{\,\,0.000000} \\
    $6$ &  & ${}^{1}{\text{S3}}$ & ${}^{1}{\text{S4}}$ & \,\,0.292770 & -0.292770 \\
    $6$ & $\hat{t}_6$ & ${}^{1}{\text{S2}}$ & ${}^{1}{\text{S3}}$ & \,\,3.558418 & -3.558417 \\
    $6$ & $\hat{t}_7$ & ${}^{1}{\text{S3}}$ & ${}^{1}{\text{S4}}$ & -2.535463 & \,\,2.535463 \\
    \noalign{\vskip 0.5ex} 
    \hline
    \noalign{\vskip 0.5ex} 
    $7$ & $\hat{f}_2$ & ${}^{2}{\text{G2}}$ & ${}^{2}{\text{G3}}$ & \,\,0.000000 & \,\,0.000001 \\
    $7$ &  & ${}^{2}{\text{G2}}$ & ${}^{2}{\text{G4}}$ & \,\,0.000000 & \,\,0.000002 \\
    $7$ &  & ${}^{2}{\text{G2}}$ & ${}^{2}{\text{G5}}$ & \,\,0.000000 & -0.000001 \\
    $7$ &  & ${}^{2}{\text{G2}}$ & ${}^{2}{\text{G6}}$ & \,\,0.000000 & -0.000003 \\
    $7$ &  & ${}^{2}{\text{G5}}$ & ${}^{2}{\text{G10}}$ & \,\,0.000000 & \,\,0.000001 \\
    $7$ &  & ${}^{2}{\text{H1}}$ & ${}^{2}{\text{H5}}$ & \,\,0.000000 & -0.000002 \\
    $7$ &  & ${}^{2}{\text{H1}}$ & ${}^{2}{\text{H7}}$ & \,\,0.000000 & \,\,0.000001 \\
    $7$ &  & ${}^{2}{\text{H2}}$ & ${}^{2}{\text{H5}}$ & \,\,0.000000 & \,\,0.000002 \\
    $7$ &  & ${}^{2}{\text{H2}}$ & ${}^{2}{\text{H7}}$ & \,\,0.000000 & \,\,0.000008 \\
    $7$ &  & ${}^{2}{\text{H5}}$ & ${}^{2}{\text{H8}}$ & \,\,0.000000 & -0.000002 \\
    $7$ &  & ${}^{2}{\text{H7}}$ & ${}^{2}{\text{H8}}$ & \,\,0.000000 & \,\,0.000003 \\
    $7$ &  & ${}^{2}{\text{H7}}$ & ${}^{2}{\text{H9}}$ & \,\,0.000000 & -0.000001 \\
    $7$ &  & ${}^{2}{\text{I1}}$ & ${}^{2}{\text{I4}}$ & \,\,0.000000 & \,\,0.000002 \\
    $7$ &  & ${}^{2}{\text{I1}}$ & ${}^{2}{\text{I5}}$ & \,\,0.000000 & \,\,0.000002 \\
    $7$ &  & ${}^{2}{\text{I3}}$ & ${}^{2}{\text{I9}}$ & \,\,0.000000 & \,\,0.000001 \\
    $7$ &  & ${}^{2}{\text{I4}}$ & ${}^{2}{\text{I6}}$ & \,\,0.000000 & \,\,0.000001 \\
    $7$ &  & ${}^{2}{\text{I4}}$ & ${}^{2}{\text{I7}}$ & \,\,0.000000 & -0.000002 \\
    $7$ &  & ${}^{2}{\text{I4}}$ & ${}^{2}{\text{I8}}$ & \,\,0.000000 & -0.000003 \\
    $7$ &  & ${}^{2}{\text{I4}}$ & ${}^{2}{\text{I9}}$ & \,\,0.000000 & \,\,0.000006 \\
    $7$ &  & ${}^{2}{\text{I5}}$ & ${}^{2}{\text{I6}}$ & \,\,0.000000 & \,\,0.000001 \\
    $7$ &  & ${}^{2}{\text{I5}}$ & ${}^{2}{\text{I8}}$ & \,\,0.000000 & -0.000001 \\
    $7$ &  & ${}^{2}{\text{I5}}$ & ${}^{2}{\text{I9}}$ & \,\,0.000000 & \,\,0.000001 \\
    $7$ &  & ${}^{2}{\text{K1}}$ & ${}^{2}{\text{K3}}$ & \,\,0.000000 & -0.000002 \\
    $7$ &  & ${}^{2}{\text{K1}}$ & ${}^{2}{\text{K4}}$ & \,\,0.000000 & -0.000002 \\
    $7$ &  & ${}^{2}{\text{K1}}$ & ${}^{2}{\text{K5}}$ & \,\,0.000000 & -0.000004 \\
    $7$ &  & ${}^{2}{\text{K2}}$ & ${}^{2}{\text{K6}}$ & \,\,0.000000 & -0.000002 \\
    $7$ &  & ${}^{2}{\text{K2}}$ & ${}^{2}{\text{K7}}$ & \,\,0.000000 & \,\,0.000001 \\
    $7$ &  & ${}^{2}{\text{K3}}$ & ${}^{2}{\text{K6}}$ & \,\,0.000000 & \,\,0.000001 \\
    $7$ &  & ${}^{2}{\text{K3}}$ & ${}^{2}{\text{K7}}$ & \,\,0.000000 & \,\,0.000001 \\
    $7$ &  & ${}^{2}{\text{K4}}$ & ${}^{2}{\text{K6}}$ & \,\,0.000000 & -0.000004 \\
    $7$ &  & ${}^{2}{\text{K4}}$ & ${}^{2}{\text{K7}}$ & \,\,0.000000 & \,\,0.000003 \\
    $7$ &  & ${}^{2}{\text{K5}}$ & ${}^{2}{\text{K6}}$ & \,\,0.000000 & \,\,0.000001 \\
    $7$ &  & ${}^{2}{\text{K5}}$ & ${}^{2}{\text{K7}}$ & \,\,0.000000 & -0.000001 \\
    $7$ &  & ${}^{2}{\text{L1}}$ & ${}^{2}{\text{L3}}$ & \,\,0.000000 & -0.000002 \\
    $7$ &  & ${}^{2}{\text{L3}}$ & ${}^{2}{\text{L5}}$ & \,\,0.000000 & -0.000001 \\
    $7$ &  & ${}^{2}{\text{M1}}$ & ${}^{2}{\text{M3}}$ & \,\,0.000000 & \,\,0.000001 \\
    $7$ &  & ${}^{2}{\text{M1}}$ & ${}^{2}{\text{M4}}$ & \,\,0.000000 & \,\,0.000001 \\
    $7$ & $\hat{f}_4$ & ${}^{2}{\text{G2}}$ & ${}^{2}{\text{G5}}$ & \,\,0.000000 & -0.000001 \\
    $7$ &  & ${}^{2}{\text{G2}}$ & ${}^{2}{\text{G6}}$ & \,\,0.000000 & \,\,0.000001 \\
    $7$ &  & ${}^{2}{\text{H1}}$ & ${}^{2}{\text{H7}}$ & \,\,0.000000 & -0.000002 \\
    $7$ &  & ${}^{2}{\text{H2}}$ & ${}^{2}{\text{H7}}$ & \,\,0.000000 & \,\,0.000003 \\
    $7$ &  & ${}^{2}{\text{H7}}$ & ${}^{2}{\text{H8}}$ & \,\,0.000000 & -0.000002 \\
    $7$ &  & ${}^{2}{\text{H7}}$ & ${}^{2}{\text{H9}}$ & \,\,0.000000 & \,\,0.000002 \\
    $7$ &  & ${}^{2}{\text{I1}}$ & ${}^{2}{\text{I4}}$ & \,\,0.000000 & -0.000001 \\
    $7$ &  & ${}^{2}{\text{I4}}$ & ${}^{2}{\text{I8}}$ & \,\,0.000000 & \,\,0.000002 \\
    $7$ &  & ${}^{2}{\text{I4}}$ & ${}^{2}{\text{I9}}$ & \,\,0.000000 & -0.000001 \\
    $7$ &  & ${}^{2}{\text{I5}}$ & ${}^{2}{\text{I6}}$ & \,\,0.000000 & -0.000001 \\
    $7$ &  & ${}^{2}{\text{I5}}$ & ${}^{2}{\text{I7}}$ & \,\,0.000000 & \,\,0.000001 \\
    $7$ &  & ${}^{2}{\text{I5}}$ & ${}^{2}{\text{I9}}$ & \,\,0.000000 & \,\,0.000001 \\
    $7$ &  & ${}^{2}{\text{K1}}$ & ${}^{2}{\text{K2}}$ & \,\,0.000000 & \,\,0.000002 \\
    $7$ &  & ${}^{2}{\text{K1}}$ & ${}^{2}{\text{K3}}$ & \,\,0.000000 & \,\,0.000001 \\
    $7$ &  & ${}^{2}{\text{K1}}$ & ${}^{2}{\text{K4}}$ & \,\,0.000000 & -0.000001 \\
    $7$ &  & ${}^{2}{\text{K1}}$ & ${}^{2}{\text{K5}}$ & \,\,0.000000 & -0.000002 \\
    $7$ &  & ${}^{2}{\text{K2}}$ & ${}^{2}{\text{K7}}$ & \,\,0.000000 & \,\,0.000001 \\
    $7$ &  & ${}^{2}{\text{K3}}$ & ${}^{2}{\text{K6}}$ & \,\,0.000000 & -0.000001 \\
    $7$ &  & ${}^{2}{\text{K4}}$ & ${}^{2}{\text{K6}}$ & \,\,0.000000 & -0.000002 \\
    $7$ &  & ${}^{2}{\text{K5}}$ & ${}^{2}{\text{K7}}$ & \,\,0.000000 & -0.000002 \\
    $7$ &  & ${}^{2}{\text{L1}}$ & ${}^{2}{\text{L3}}$ & \,\,0.000000 & \,\,0.000001 \\
    $7$ &  & ${}^{2}{\text{L3}}$ & ${}^{2}{\text{L5}}$ & \,\,0.000000 & -0.000001 \\
    $7$ &  & ${}^{2}{\text{M1}}$ & ${}^{2}{\text{M3}}$ & \,\,0.000000 & -0.000001 \\
    $7$ &  & ${}^{2}{\text{M1}}$ & ${}^{2}{\text{M4}}$ & \,\,0.000000 & -0.000001 \\
    $7$ & $\hat{f}_6$ & ${}^{2}{\text{G2}}$ & ${}^{2}{\text{G4}}$ & \,\,0.000000 & -0.000001 \\
    $7$ &  & ${}^{2}{\text{G2}}$ & ${}^{2}{\text{G5}}$ & \,\,0.000000 & \,\,0.000003 \\
    $7$ &  & ${}^{2}{\text{G2}}$ & ${}^{2}{\text{G6}}$ & \,\,0.000000 & -0.000001 \\
    $7$ &  & ${}^{2}{\text{H1}}$ & ${}^{2}{\text{H7}}$ & \,\,0.000000 & \,\,0.000003 \\
    $7$ &  & ${}^{2}{\text{H2}}$ & ${}^{2}{\text{H5}}$ & \,\,0.000000 & \,\,0.000001 \\
    $7$ &  & ${}^{2}{\text{H2}}$ & ${}^{2}{\text{H7}}$ & \,\,0.000000 & -0.000001 \\
    $7$ &  & ${}^{2}{\text{H5}}$ & ${}^{2}{\text{H8}}$ & \,\,0.000000 & \,\,0.000001 \\
    $7$ &  & ${}^{2}{\text{H5}}$ & ${}^{2}{\text{H9}}$ & \,\,0.000000 & -0.000001 \\
    $7$ &  & ${}^{2}{\text{H7}}$ & ${}^{2}{\text{H8}}$ & \,\,0.000000 & -0.000001 \\
    $7$ &  & ${}^{2}{\text{H7}}$ & ${}^{2}{\text{H9}}$ & \,\,0.000000 & \,\,0.000004 \\
    $7$ &  & ${}^{2}{\text{I1}}$ & ${}^{2}{\text{I5}}$ & \,\,0.000000 & -0.000001 \\
    $7$ &  & ${}^{2}{\text{I5}}$ & ${}^{2}{\text{I6}}$ & \,\,0.000000 & \,\,0.000001 \\
    $7$ &  & ${}^{2}{\text{I5}}$ & ${}^{2}{\text{I7}}$ & \,\,0.000000 & \,\,0.000001 \\
    $7$ &  & ${}^{2}{\text{I5}}$ & ${}^{2}{\text{I9}}$ & \,\,0.000000 & \,\,0.000001 \\
    $7$ &  & ${}^{2}{\text{K1}}$ & ${}^{2}{\text{K2}}$ & \,\,0.000000 & -0.000002 \\
    $7$ &  & ${}^{2}{\text{K1}}$ & ${}^{2}{\text{K4}}$ & \,\,0.000000 & \,\,0.000001 \\
    $7$ &  & ${}^{2}{\text{K1}}$ & ${}^{2}{\text{K5}}$ & \,\,0.000000 & \,\,0.000003 \\
    $7$ &  & ${}^{2}{\text{K2}}$ & ${}^{2}{\text{K6}}$ & \,\,0.000000 & \,\,0.000001 \\
    $7$ &  & ${}^{2}{\text{K2}}$ & ${}^{2}{\text{K7}}$ & \,\,0.000000 & -0.000001 \\
    $7$ &  & ${}^{2}{\text{K5}}$ & ${}^{2}{\text{K6}}$ & \,\,0.000000 & -0.000001 \\
    $7$ &  & ${}^{2}{\text{K5}}$ & ${}^{2}{\text{K7}}$ & \,\,0.000000 & -0.000001 \\
    $7$ &  & ${}^{2}{\text{K6}}$ & ${}^{2}{\text{K6}}$ & \,\,0.000634 & \,\,0.000638 \\
    $7$ &  & ${}^{2}{\text{M1}}$ & ${}^{2}{\text{M4}}$ & \,\,0.000000 & -0.000002 \\
    $7$ & $\hat{t}_2$ & ${}^{2}{\text{F2}}$ & ${}^{2}{\text{F3}}$ & \,\,0.235970 & \,\,0.196756 \\
    $7$ &  & ${}^{2}{\text{F2}}$ & ${}^{2}{\text{F8}}$ & -0.410326 & -0.359701 \\
    $7$ &  & ${}^{2}{\text{F3}}$ & ${}^{2}{\text{F4}}$ & -0.793039 & -0.813518 \\
    $7$ &  & ${}^{2}{\text{F3}}$ & ${}^{2}{\text{F6}}$ & -0.699854 & -0.635095 \\
    $7$ &  & ${}^{2}{\text{F4}}$ & ${}^{2}{\text{F8}}$ & \,\,0.250000 & \,\,0.276438 \\
    $7$ &  & ${}^{2}{\text{F6}}$ & ${}^{2}{\text{F8}}$ & \,\,0.000000 & -0.083604 \\
    \textcolor{blue}{$7$} & \textcolor{blue}{$\hat{t}_3$} & \textcolor{blue}{${}^{2}{\text{F1}}$} & \textcolor{blue}{${}^{2}{\text{F8}}$} & \textcolor{blue}{-4.535574} & \textcolor{blue}{-3.239695} \\
    $7$ & $\hat{t}_4$ & ${}^{2}{\text{F3}}$ & ${}^{2}{\text{F10}}$ & -0.358569 & -0.320150 \\
    $7$ &  & ${}^{2}{\text{F3}}$ & ${}^{2}{\text{F9}}$ & \,\,0.478091 & \,\,0.490897 \\
    $7$ &  & ${}^{2}{\text{F5}}$ & ${}^{2}{\text{F8}}$ & -0.377964 & -0.357716 \\
    $7$ &  & ${}^{2}{\text{F7}}$ & ${}^{2}{\text{F8}}$ & \,\,0.000000 & \,\,0.035071 \\
    $7$ & $\hat{t}_6$ & ${}^{2}{\text{F2}}$ & ${}^{2}{\text{F3}}$ & -1.452718 & -1.141421 \\
    $7$ &  & ${}^{2}{\text{F4}}$ & ${}^{2}{\text{F8}}$ & \,\,0.000000 & -0.209876 \\
    $7$ &  & ${}^{2}{\text{F6}}$ & ${}^{2}{\text{F8}}$ & -2.725541 & -2.061853 \\
    $7$ & $\hat{t}_7$ & ${}^{2}{\text{F3}}$ & ${}^{2}{\text{F10}}$ & \,\,1.725164 & \,\,1.392453 \\
    $7$ &  & ${}^{2}{\text{F3}}$ & ${}^{2}{\text{F9}}$ & \,\,0.000000 & -0.110903 \\
    $7$ &  & ${}^{2}{\text{F5}}$ & ${}^{2}{\text{F8}}$ & \,\,0.000000 & -0.175354 \\
    $7$ &  & ${}^{2}{\text{F7}}$ & ${}^{2}{\text{F8}}$ & \,\,1.889822 & \,\,1.586100 \\
    \noalign{\vskip 0.5ex} 
    \hline
    \noalign{\vskip 0.5ex}
    \textcolor{red}{$8$} & 
    \textcolor{red}{$\hat{t}_3$} &
    \textcolor{red}{$\LSterm{1}{S1}$} &
    \textcolor{red}{$\LSterm{1}{S3}$} &
    \textcolor{red}{\,\,-5.7371} & \textcolor{red}{5.7371} \\
    \textcolor{red}{$8$} & 
    \textcolor{red}{$\hat{t}_6$} &
    \textcolor{red}{$\LSterm{1}{S2}$} &
    \textcolor{red}{$\LSterm{1}{S3}$} &
    \textcolor{red}{\,\,-3.55842} & \textcolor{red}{3.55842} \\
    \textcolor{red}{$8$} & 
    \textcolor{red}{$\hat{t}_7$} &
    \textcolor{red}{$\LSterm{1}{S3}$} &
    \textcolor{red}{$\LSterm{1}{S4}$} &
    \textcolor{red}{\,\,2.53546} & \textcolor{red}{-2.53546} \\
    \textcolor{red}{$8$} & 
    \textcolor{red}{$\hat{t}_2$} &
    \textcolor{red}{$\LSterm{1}{S2}$} &
    \textcolor{red}{$\LSterm{1}{S3}$} &
    \textcolor{red}{\,\,-0.589802} & \textcolor{red}{-1.4863} \\
    \textcolor{red}{$8$} & 
    \textcolor{red}{$\hat{t}_4$} &
    \textcolor{red}{$\LSterm{1}{Q}$} &
    \textcolor{red}{$\LSterm{1}{Q}$} &
    \textcolor{red}{\,\,0.856893} & \textcolor{red}{0.0} \\
    \textcolor{red}{$8$} & 
    \textcolor{red}{$\hat{t}_4$} &
    \textcolor{red}{$\LSterm{1}{S3}$} &
    \textcolor{red}{$\LSterm{1}{S4}$} &
    \textcolor{red}{\,\,-0.29277} & \textcolor{red}{0.29277} \\
    \noalign{\vskip 0.5ex} 
    \hline
    \noalign{\vskip 0.5ex}
    \textcolor{red}{$12$} & \textcolor{red}{$\hat{t}_2$} & \textcolor{red}{${}^{1}{\text{G}}$} & \textcolor{red}{${}^{1}{\text{G}}$} & \textcolor{red}{\,\,-0.404061} & \textcolor{red}{\,\,0.000000} \\
    \hline
\end{supertabular}
\label{table:extra-errors}

\normalsize
\newpage

\end{center}

\begin{widetext}
\newpage
\section{Typographical errors}
\label{sec: Carnall tables}

\begin{table}[!htbp]
\begin{center}
\begin{tabular}{|c|c|c|c|c|c|}
\hline
Ln & SLJ State                 & Obsd. (cm$^{-1}$) & Calc (cm$^{-1}$)  & O-C & Comment                                            \\ \hline\hline
Nd & 4I15/2                    & 6323       & 6113         & 10       & Incompatible O-C value.
\\ \hline
Sm & 4L17/2,6P7/2              & 26859      & 26349        & 10       & Incompatible O-C value. 
\\  \hline
Sm & 4F7/2, 2N19/2             &  -         & (35612-35823)     &   -       & Number of levels in interval not given.      
\\  \hline
Sm & 4P1/2, 2H9/2, 2F5/2             &  -         & (49581-49865)     &   -       & Number of levels in interval not given.      
\\  \hline
Tb & 7F4                       &  -        & 3506          &   -    & Duplicate, the 7F4 multiplet should only have 9 states. 
\\ \hline 
Tb & 5I7                       & 36348     & 36330         &  -         & State is displaced, given energy breaks ascending order. 
\\ \hline 
Dy & 6P3/2                     & 30879      & 30062        & 17       & Incompatible O-C value.  
\\  \hline
\end{tabular}
\end{center}
\caption{Typographical errors in the energy levels used in \bill.}
\end{table}
\end{widetext}

\begin{widetext}
\section{Schematic mapping between Hamiltonains}
\label{sec: mapping}

\begin{center}
\begin{table}[!htbp]
\begin{tabular}{|c|c|c|}
\hline 
Hamiltonian & Parametric Hamiltonian & Origin\tabularnewline
\hline 
\hline 
${\cal H}_{K}+{\cal H}_{n}$ & ${\cal H}_{0}$ & single configuration\tabularnewline
\hline 
${\cal H}_{e\text{-}e}$ & $F^{(k)},\quad k=0,2,4,6$ & single configuration\tabularnewline
\hline 
${\cal H}_{e\text{-}e}$ & $\alpha,\beta,\gamma$ + screening $F^{(k)}$ & configuration interaction via $\left\langle \psi\right|G\left|m\right\rangle \left\langle m\right|G\left|\psi'\right\rangle /\Delta E_{m}$ \tabularnewline
\hline 
${\cal H}_{e\text{-}e}$ & $T^{(k)},\quad k=3,4,6,7,8$ & same as the above, for the cases that $\left|\psi\right\rangle $
and $\left|m\right\rangle $ differ in one electron\tabularnewline
\hline 
${\cal H}_{e\text{-}e}$ & $T^{(k)},\quad k=11,\ldots,19$ & configuration interaction, third order perturbation theory\tabularnewline
\hline 
${\cal H}_{\text{s-o}}$ & $\zeta$ & single configuration\tabularnewline
\hline 
${\cal H}_{e\text{-}e}+{\cal H}_{\text{s-o}}$ & $P^{(k)}$ + $M^{(k)}$ + screening $\zeta$ & configuration interaction via $\left\langle \psi\right|\Lambda\left|m\right\rangle \left\langle m\right|G\left|\psi'\right\rangle /\Delta E_{m}$ \tabularnewline
\hline 
${\cal H}_{\text{s-s}}$ & $M^{(k)},\quad k=0,2,4$ & single configuration\tabularnewline
\hline 
${\cal H}_{\text{s-o-o}}$ & $M^{(k)}$ + screening $\zeta$ & single configuration\tabularnewline

\hline 
${\cal H}_{\text{o-o}}$ & $\alpha,\beta,\gamma$ & single configuration\tabularnewline
\hline 
${\cal H}_{\text{CF}}$ & ${\cal B}_{q}^{(k)},\quad k=2,4,6,\quad q=-k,\ldots,k$ & single configuration\tabularnewline
\hline 
\end{tabular}
\caption{A summary of the mapping from the Hamiltonian in Eq. (\ref{eq: Hamiltonian contributions}) to the parametric
Hamiltonian in Eq. (\ref{eqn:hamiltonian}). }
\label{tbl: mapping}
\end{table}
\end{center}
\end{widetext}

\section{Relation between orthogonal and nonorthogonal Hamiltonians}
\label{app: orthogonalization}

In the set of Eqs. (\ref{eq: orthogonal E1})-(\ref{eq: nonorthogonal E3}) we provide relations between parameters
for Hamiltonians in the orthogonal and nonorthogonal forms. Here, we
elaborate on the calculations that led to those relations.

We start by defining the Hamiltonians in both forms, orthogonal,
\begin{align}
{\cal H}^{\text{ortho}}= & E'^{(0)} \hat{e}'_{0}+E'^{(1)} \hat{e}'_{1}+E'^{(3)} \hat{e}'_{3}\nonumber \\
 & +\alpha' \hat{e}'_{\alpha} +\beta' \hat{e}'_{\beta} +\gamma' \hat{e}'_{\gamma} +T'^{(2)} \hat{t}'_{2},
\end{align}
(for shortness, here we use the $'$ symbol instead of $\perp$, as in the main text, for representing the orthogonal operators and parameters) and non-orthogonal,
\begin{align}
{\cal H}^{\text{non}}= & E^{(0)}\hat{e}{}_{0}+E{}^{(1)}\hat{e}{}_{1}+E{}^{(3)}\hat{e}{}_{3}\nonumber \\
 & +\alpha\hat{e}{}_{\alpha}+\beta\hat{e}{}_{\beta}+\gamma\hat{e}{}_{\gamma}+T{}^{(2)}\hat{t}{}_{2},
\end{align}
in which we have only selected the elements that matter for current
discussion, and not the full parametric Hamiltonian as written in Eq.
(\ref{eqn:hamiltonian}) in the main text. Furthermore, Eq. (\ref{eqn:hamiltonian}) is written in terms of
$\hat{f}_{0},\hat{f}_{2},\hat{f}_{4},\hat{f}_{6}$ operators instead
of $\hat{e}_{0},\hat{e}_{1},\hat{e}_{3}$; however, there is a straight
forward relation between them that will be provided below. Finally,
by comparison with Eq. (\ref{eqn:hamiltonian}), notice that 
\begin{equation}
\hat{e}{}_{\alpha}=\hat{L}^{2},\;\hat{e}{}_{\beta}=\hat{{\cal C}}\left({\cal G}_{2}\right),\;\hat{e}{}_{\gamma}=\hat{{\cal C}}\left({\cal SO}_{7}\right). \label{eq: nonortho alpha beta gamma operators}
\end{equation}
The relations between orthogonal and nonorthogonal operators are provided
in Ref. \cite{Judd_Crosswhite_84},
that we replicate here for convenience,
\begin{align}
\hat{e}'_{0} & =\hat{e}_{0}\\
\hat{e}'_{1} & =\hat{e}_{1}-\frac{9\hat{e}_{0}}{13},\\
\hat{e}'_{3} & =\hat{e}_{3}\\
\hat{e}'_{\alpha} & =\frac{\hat{e}_{3}}{2}+\frac{5\hat{L}^{2}}{4}-30\hat{{\cal C}}\left({\cal G}_{2}\right),\\
\hat{e}'_{\beta} & =5\hat{{\cal C}}\left({\cal SO}_{7}\right)-6\hat{{\cal C}}\left({\cal G}_{2}\right),\\
\hat{e}'_{\gamma} & =\frac{25\hat{{\cal C}}\left({\cal SO}_{7}\right)}{2}-\frac{15N}{2}+\frac{3\hat{e}_{0}}{2}-\frac{\hat{e}_{1}}{2},\\
\hat{t}'_{2} & =\hat{t}_{2}-\frac{\left(N-2\right)\hat{e}_{3}}{70\sqrt{2}},
\end{align}
where $N$ is the number of electrons in the $f$ shell. Next, we replace
those definitions into ${\cal H}^{\text{ortho}}$ and rearrange the
terms to obtain
\begin{align}
{\cal H}^{\text{ortho}}= & \left(E'^{(0)}-\frac{9E'^{(1)}}{13}+\frac{3\gamma'}{2}\right)\hat{e}_{0}+\left(E'^{(1)}-\frac{\gamma'}{2}\right)\hat{e}_{1}\nonumber \\
 & +\left(E'^{(3)}+\frac{\alpha'}{2}-\frac{T'^{(2)}\left(N-2\right)}{70\sqrt{2}}\right)\hat{e}_{3}\nonumber \\
 & +\frac{5\alpha'}{4}\hat{L}^{2}-\left(\beta'6+\alpha'30\right)\hat{{\cal C}}\left({\cal G}_{2}\right)\nonumber \\
 & +\left(\gamma'\frac{25}{2}+\beta'5\right)\hat{{\cal C}}\left({\cal SO}_{7}\right)-\gamma'\frac{15N}{2}+T'^{(2)}\hat{t}_{2}.
\end{align}
Therefore, comparing the coefficients above with nonorthogonal Hamiltonian
${\cal H}^{\text{non}}$, acknowledging Eqs. (\ref{eq: nonortho alpha beta gamma operators}),
both Hamiltonian become the same (disregarding the constant term)
for the following relations
\begin{align}
E^{(0)} & =E'^{(0)}-\frac{9E'^{(1)}}{13}+\frac{3\gamma'}{2},\label{eq: E0 set of equations}\\
E^{(1)} & =E'^{(1)}-\frac{\gamma'}{2},\\
E^{(3)} & =E'^{(3)}+\frac{\alpha'}{2}-\frac{\left(N-2\right)T'^{(2)}}{70\sqrt{2}},\\
\alpha & =\frac{5\alpha'}{4},\\
\beta & =-\left(6\beta'+30\alpha'\right),\\
\gamma & =\frac{25}{2}\gamma'+5\beta',\\
T^{(2)} & =T'^{(2)}. \label{eq: T2 set of equations}
\end{align}
Inverting the identities above for $\alpha'$, $\beta'$, $\gamma'$,
and $T'^{(2)}$ we find the relations in Eqs. (\ref{eq: orthogonal E1})-(\ref{eq: T2 prime}) in the main text. 
Now, we are left with the task of relating $E$'s and $F$'s.
However, those are well explained in chap. 2 sec. 9 of Ref. \cite{wybourne1965spectroscopic},
that kindly provides a direct relation between coefficients, 
as already shown in Eqs. (\ref{eq: nonorthogonal E1})-(\ref{eq: nonorthogonal E3}) in the main text.
 
\begin{center}
\begin{table*}
    \includegraphics[width=0.99\textwidth]{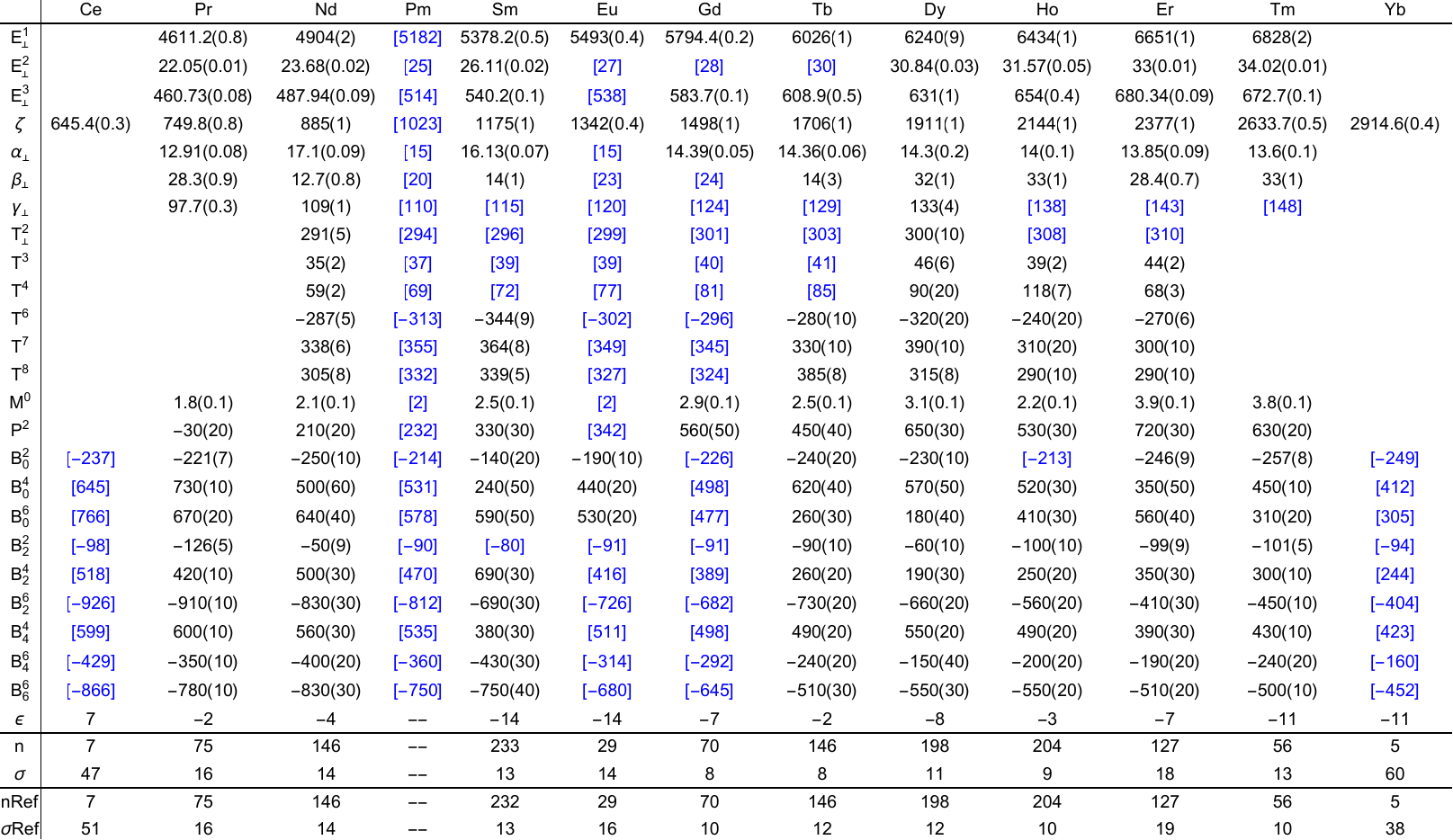}
    \caption{Fitted model parameters using \qlanth with orthogonal operators. Parameters in brackets were held fixed during fitting. In the case of Eu, Gd, and Tb, some $\text{E}^i_\perp$ parameters were constrained according to $\text{E}^2_\perp = 0.0049 \text{E}^1_\perp, \text{and/or }  
    \text{E}^3_\perp = 0.098 \text{E}^1_\perp$. The parameters for Pm are interpolated. $\sigma_\text{REF}$ and $n_\text{REF}$ are from \bill. The data was fitted in the following order: Pr, Nd, Dy, Ce, Sm, Ho, Er, Tm, Yb, Tb, Eu, Gd.}
       \label{table:qlanthLaF3-mostly-orthogonal}
\end{table*}
\end{center}

We emphasize that the transformation above is not the complete story, because for a completely orthogonal parametric Hamiltonian, we would still need to orthogonalize the operators related to the $M$'s and $P$'s coefficients. This lingering part of the Hamiltonian may be made orthogonal using the $\hat{z}_i$ operators initially described by Judd, Crosswhite, and Crosswhite in their description of the intra-atomic magnetic interactions \cite{judd_intra-atomic_1968} and explained in Ref. \cite{Judd-2008-JPB}. However, we will not consider it here, as it adds additional parameters that are not yet well explored in the literature. Having said that, the full (mostly) orthogonal Hamiltonian considered here is parametrized as
\begin{align}
    {\cal H}^{\text{ortho}}_{\text{para}} &=  \ham_{0}
     + \sum_{k=0,2,4,6} E'^{(k)} \hat{e}'_{k}
     + \spinZeta \sum_{i=1}^N  \paren{ \hat{s}_i \cdot \hat{l}_i} \nonumber \\
     &\,\,  + \alpha' \hat{e}'_{\alpha}
      + \beta' \hat{e}'_{\beta}
       + \gamma' \hat{e}'_{\gamma} \nonumber \\
     &\,\,\,\,  + {T'^{(2)}} \hat{t}'_2 
      + \hamEffectiveThreeBody \nonumber \\
     &\,\,\,\,\,\, 
     + \sum_{k=2,4,6} P^{(k)} \hat{p}_k + \sum_{k=0,2,4} M^{(k)} \hat{m}_k \nonumber \\
     &\,\,\,\,\,\,\,\,  + \hamCrystalFieldALT.
    \label{eqn:mostly-orthogonal-hamiltonian}
\end{align}

Table \ref{table:qlanthLaF3-mostly-orthogonal} shows the fitted parameters in ${\cal H}^{\text{ortho}}_{\text{para}}$ (notice the equivalence in notations $(') \equiv \, \perp$). The fitting procedure used the same sequential manner as described in the main paper, i.e., buidling up regression models for the parameters that are held fixed while fitting. Contrasting the orthogonal and nonorthogonal approaches against \bill, Table \ref{table:qlanthLaF3} in the main text shows $\sigma$ to be smaller than those values quoted by Carnall \textit{et al.}  in 2 out of 12 cases. For the approach with the orthogonal operator basis shown here, in Table \ref{table:qlanthLaF3-mostly-orthogonal} we found $\sigma$ to be smaller in 7 out of 12 cases. That shows the expected tendency of the orthogonal approach leading to better fittings.

Note that the parameters for Ce, Pr, Nd, and Dy in Tables \ref{table:qlanthLaF3} and \ref{table:qlanthLaF3-mostly-orthogonal} seem to coincide, including the orthogonalized parameters, as they are related by Eqs. (\ref{eq: orthogonal E1})-(\ref{eq: nonorthogonal E3}) in the main text. That reflects the deterministic approach taken here and described in the main text. The fitting initial conditions are the same in both approaches, leading to a similar converged values, although not exactly equal within the numerical precision. 
This fact changes for Sm (the next element in the fitting), because a regression made for fixing $\gamma$ (non-orthogonal) and $\gamma_\perp$ (orthogonal) lead to different results. The same is true for the subsequent fittings.
Finally, we also compared the relative uncertainties in both tables, and unexpectedly, we found that the orthogonal basis leads to larger uncertainty values on average (9 out of 12 cases).

\section{Global fitting analysis}
\label{app: global fitting}

In the main text we employed the same Levenberg-Marquardt method used in Ref. \cite{carnall_systematic_1989} to obtain the fitted parameters. In order to test whether or not we have found a global minimum, we decided to try a stochastic optimization algorithm.
In particular, we used the method Adam\cite{kingma2017adam} from the PyTorch library \cite{paszke2019pytorch}.
Note that \qlanth has the feature of exporting the Hamiltonian to Python.
We exemplify the point by finding new parameters for $\text{Er}^{3+}\text{:LaF}_3$.
We always used the values in Ref. \cite{carnall_systematic_1989} as initial guess, among the many attempts performed.
We avoided exploring parameter values too far from the original guess, because we know it has physical significance and serves as a meaningful reference point.
This is controlled by the learning rate ($lr$) parameter, that we typically set $lr=0.1$.
We checked that $2000$ iterations usually led to a good convergence, although we made attempts up to $50000$ iterations.
Our code is available at Ref. \cite{optimizationTharnier}.
As a result, we could bring the average agreement to the experimental energy levels ($\sigma$, as described in the main text) down to $\sigma = 15 \text{ cm}^{-1}$, which is a $20\%$ improvement over the value quoted in Ref. \cite{carnall_systematic_1989}.
The set of new parameters are listed in Table \ref{tbl: Er-LaF3}.

\begin{center}
\begin{table}[!htbp]
\begin{tabular}{|c|c||c|c|}
\hline 
Parameter & Value & Parameter & Value\tabularnewline
\hline 
\hline 
$F^{(2)}$ & $97592(71)$ & $T^{(6)}$ & $-289(23)$\tabularnewline
\hline 
$F^{(4)}$ & $68036(142)$ & $T^{(7)}$ & $336(38)$\tabularnewline
\hline 
$F^{(6)}$ & $54187(136)$ & $T^{(8)}$ & $364(39)$\tabularnewline
\hline 
$\zeta$ & $2377(5)$ & $B_{0}^{(2)}$ & $-242(33)$\tabularnewline
\hline 
$\alpha$ & $17.3(0.5)$ & $B_{0}^{(4)}$ & $367(238)$\tabularnewline
\hline 
$\beta$ & $-583(21)$ & $B_{0}^{(6)}$ & $522(191)$\tabularnewline
\hline 
$\gamma$ & $\left[1800\right]$ & $B_{2}^{(2)}$ & $-96(29)$\tabularnewline
\hline 
$M^{(0)}$ & $3.8(0.4)$ & $B_{2}^{(4)}$ & $357(126)$\tabularnewline
\hline 
$P^{(2)}$ & $700(141)$ & $B_{2}^{(6)}$ & $-424(142)$\tabularnewline
\hline 
$T^{(2)}$ & $\left[400\right]$ & $B_{4}^{(4)}$ & $417(115)$\tabularnewline
\hline 
$T^{(3)}$ & $45(10)$ & $B_{4}^{(6)}$ & $-203(85)$\tabularnewline
\hline 
$T^{(4)}$ & $67(12)$ & $B_{6}^{(6)}$ & $-513(111)$\tabularnewline
\hline 
\end{tabular}
\caption{Fitted parameters for $\trion{Er}\text{:LaF}_{3}$ using Adam method from
PyTorch Library. Values with their uncertainties given in $\text{cm}^{-1}$.
Values for $\gamma$ and $T^{(2)}$ were held fixed during fitting.}
\label{tbl: Er-LaF3}
\end{table}
\end{center}

\section{$\text{Ln}^{3+}\!$ in \liyorite}
\label{appendix:liyorite}

A fitting procedure identical to the one described in the main text for \LaF was applied to the description of lanthanide ions in \liyorite. This is an important prospective crystal host of quantum applications\cite{Beckert:2024aa}. The work of \cheng was used in this case as a comparison point to our results. The experimental references cited by them were collated, and the data fitted using a crystal field of D2d symmetry. The results of fitting are shown in Table \ref{table:qlanth-LiYF4}. Note that our fits include all experimental data, including some that Cheng et al. must have chosen to omit given the differences between $\text{n}$ and $\text{n}_\text{REF}$. Nevertheless, the resulting quality of our fits, as measured by $\sigma$, was improved in 6 out of 11 cases, and was equal or slightly worse in 5 out of 11 cases.

\begin{widetext}

\begin{center}
\begin{table*}[b]
    \includegraphics[width=0.99\textwidth]{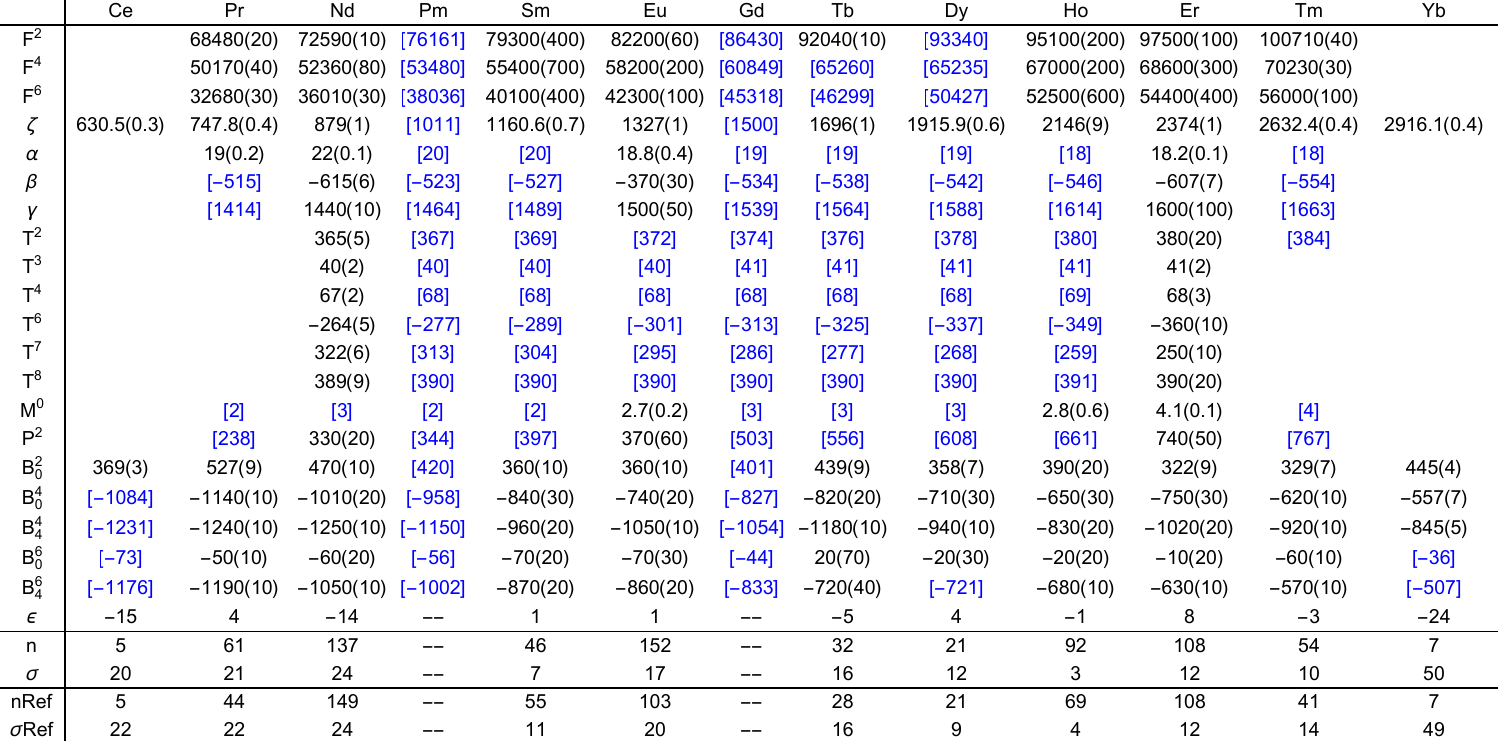}
    \caption{Fitted model parameters for \liyorite using \qlanth. Following \cheng, parameters in brackets were held fixed during fitting. In the case of Tb $\Fk{k}$ parameters were constrained according to $\text{F}^4 = 0.709 \text{F}^2$ and $\text{F}^6 = 0.503 \text{F}^2$. Data was fitted in the following order: Er, Nd, Eu, Ho, Sm, Pr, Tm, Yb, Ce, Tb, Dy, Gd, Pm. There was no data for Pm and Gd, as such, the parameters shown are those interpolated using the parameters for the other ions. $\text{n}_\text{REF}$ and $\sigma_\text{REF}$ are from \cheng.}
\label{table:qlanth-LiYF4}
\end{table*}
\end{center}
\end{widetext}

\end{document}